\documentclass[preprint,tightenlines,nofootinbib,showpacs,amsmath,amssymb,byrevtex,floatfix]{revtex4}

\usepackage{graphicx,comment}
\usepackage{amssymb,latexsym}
\usepackage{amsmath,amsbsy}

\begin{document}

\def\OMIT#1{{}}
\newcommand{\gsim}{\raisebox{-0.7ex}{$\stackrel{\textstyle >}{\sim}$ }}
\newcommand{\lsim}{\raisebox{-0.7ex}{$\stackrel{\textstyle <}{\sim}$ }}

% Field definitions
\def\Bbar{\overline{B}}
\def\bbar{\overline{b}}
\def\Tbar{\overline{T}}
\def\tbar{\overline{t}}
\def\cB{{\mathcal B}}
\def\cT{{\mathcal T}}
\def\cBbar{\overline{\cal B}}
\def\cTbar{\overline{\cal T}}
\def\cM{\mathcal M}
\def\cA{\mathcal A}

%coefficients
\def\am{\alpha_M}
\def\bm{\beta_M}
\def\gm{\gamma_M}
\def\smb{\sigma_M}
\def\smt{\overline{\sigma}_M}
\def\C{{\mathcal C}}
\def\tb{{\tilde b}}

% Greek letters
\def\a{{\alpha}}
\def\b{{\beta}}
\def\D{{\Delta}}
\def\d{{\delta}}
\def\e{{\epsilon}}
\def\g{{\gamma}}
\def\G{{\Gamma}}
\def\k{{\kappa}}
\def\l{{\lambda}}
\def\L{{\Lambda}}
\def\o{{\omega}}
\def\O{{\Omega}}
\def\S{{\Sigma}}
\def\s{{\sigma}}
\def\th{{\theta}}
\def\X{{\Xi}}

% quark and meson masses
\def\mq{m_q}
\def\mbar{\overline{m}}
\def\mup{m_u}
\def\md{m_d}
\def\ms{m_s}
\def\mj{m_j}
\def\ml{m_l}
\def\mr{m_r}

\def\mphi{m_\phi}
\def\mp{m_\pi}
\def\mk{m_K}
\def\me{m_\eta}
\def\mes{m_{\eta_s}}
\def\mju{m_{ju}}
\def\mjj{m_{\eta_j}}
\def\mjs{m_{js}}
\def\mjr{m_{jr}}
\def\mru{m_{ru}}
\def\mrs{m_{rs}}
\def\mrr{m_{\eta_r}}
\def\mX{m_X}

% Baryon Masses
%%%%% Octet
\def\MN{M_N^{(1)}}
\def\MS{M_\S^{(1)}}
\def\ML{M_\L^{(1)}}
\def\MX{M_\X^{(1)}}
%%%%% Decuplet
\def\MD{M_\D^{(1)}}
\def\MSS{M_{\S^{*}}^{(1)}}
\def\MXS{M_{\X^{*}}^{(1)}}
\def\MO{M_{\O^-}^{(1)}}
%
% Sea Baryons
%%%%% Octet
\def\MsSj{M_{\S_j}^{(1)}}
\def\MsSr{M_{\S_r}^{(1)}}
\def\MsXsj{M_{\X^{(6)}_j}^{(1)}}
\def\MsXsr{M_{\X^{(6)}_r}^{(1)}}
\def\MsXtj{M_{\X^{({\overline 3})}_j}^{(1)}}
\def\MsXtr{M_{\X^{({\overline 3})}_r}^{(1)}}
\def\MsLj{M_{\L_j}^{(1)}}
\def\MsLr{M_{\L_r}^{(1)}}
\def\MsOj{M_{\O_j}^{(1)}}
\def\MsOr{M_{\O_r}^{(1)}}
%%%%% Decuplet
\def\MsSSj{M_{\S^{*}_j}^{(1)}}
\def\MsSSr{M_{\S^{*}_r}^{(1)}}
\def\MsXSj{M_{\X^{*}_j}^{(1)}}
\def\MsXSr{M_{\X^{*}_r}^{(1)}}
\def\MsOmj{M_{\O^{*}_j}^{(1)}}
\def\MsOmr{M_{\O^{*}_r}^{(1)}}
%
% Ghost Baryons
%%%%% Octet
\def\MgSu{M_{{\tilde \S}_{\tilde u}}^{(1)}}
\def\MgSs{M_{{\tilde \S}_{\tilde s}}^{(1)}}
\def\MgXsu{M_{{\tilde \X}^{(6)}_{\tilde u}}^{(1)}}
\def\MgXss{M_{{\tilde \X}^{(6)}_{\tilde s}}^{(1)}}
\def\MgXtu{M_{{\tilde \X}^{({\overline 3})}_{\tilde u}}^{(1)}}
\def\MgXts{M_{{\tilde \X}^{({\overline 3})}_{\tilde s}}^{(1)}}
\def\MgLu{M_{{\tilde \L}_{\tilde u}}^{(1)}}
\def\MgLs{M_{{\tilde \L}_{\tilde s}}^{(1)}}
\def\MgOu{M_{{\tilde \O}_{\tilde u}}^{(1)}}
\def\MgOs{M_{{\tilde \O}_{\tilde s}}^{(1)}}
%%%%% Decuplet
\def\MgSSu{M_{{\tilde \S}^{*}_{\tilde u}}^{(1)}}
\def\MgSSs{M_{{\tilde \S}^{*}_{\tilde s}}^{(1)}}
\def\MgXSu{M_{{\tilde \X}^{*}_{\tilde u}}^{(1)}}
\def\MgXSs{M_{{\tilde \X}^{*}_{\tilde s}}^{(1)}}
\def\MgOmu{M_{{\tilde \O}^{*}_{\tilde u}}^{(1)}}
\def\MgOms{M_{{\tilde \O}^{*}_{\tilde s}}^{(1)}}

% useful math
\def\cL{\mathcal L}
\def\hL{\mathfrak L}
\def\ol#1{\overline{#1}}
\def\Dslash{D\hskip-0.65em /}
\def\dslash{\partial \hskip-0.5em /}
\def\vD{\vit\cdot D}
\def\vslash{v\hskip-0.50em /}
\def\vslashh{v\hskip-0.35em /}
\def\vit{{\it v}}
\def\Or{{\cal O}}
\def\ln{{\rm log}}

\def\Lphi{\cL_\phi}
\def\Ltphi{{\overline \cL}_\phi}
\def\Lpi{\cL_\pi}
\def\Ltpi{{\overline \cL}_\pi}
\def\LK{\cL_K}
\def\LtK{{\overline \cL}_K}
\def\Leta{\cL_\eta}
\def\Lteta{{\overline \cL}_\eta}
\def\Lju{\cL_{ju}}
\def\Ltju{{\overline \cL}_{ju}}
\def\Ljs{\cL_{js}}
\def\Ltjs{{\overline \cL}_{js}}
\def\Lru{\cL_{ru}}
\def\Ltru{{\overline \cL}_{ru}}
\def\Lrs{\cL_{rs}}
\def\Ltrs{{\overline \cL}_{rs}}
\def\Ljj{\cL_{\eta_j}}
\def\Ltjj{{\overline \cL}_{jj}}
\def\Ljr{\cL_{jr}}
\def\Ltjr{{\overline \cL}_{jr}}
\def\Lrr{\cL_{\eta_r}}
\def\Ltrr{{\overline \cL}_{rr}}
\def\Lss{\cL_{\eta_s}}
\def\Ltss{{\overline \cL}_{ss}}
\def\LX{\cL_X}
\def\LtX{{\overline \cL}_X}

\def\Fphiphi{\cF_{\phi,{\phi^\prime}}}
\def\Fpp{\cF_{\pi,\pi}}
\def\Fps{\cF_{\pi,\eta_s}}
\def\Fss{\cF_{\eta_s,\eta_s}}

\def\cF{\mathcal F}
\def\Fphi{\cF_\phi}
\def\Fpi{\cF_\pi}
\def\FK{\cF_K}
\def\Feta{\cF_\eta}
\def\Fju{\cF_{ju}}
\def\Fru{\cF_{ru}}
\def\Fjs{\cF_{js}}
\def\Frs{\cF_{rs}}
\def\Fs{\cF_{\eta_s}}
\def\FX{\cF_X}

\def\cJ{\mathcal J}
\def\Jphi{\cJ_\phi}
\def\Jpi{\cJ_\pi}
\def\JK{\cJ_K}
\def\Jeta{\cJ_\eta}
\def\Jss{\cJ_{\eta_s}}
\def\Jju{\cJ_{ju}}
\def\Jjj{\cJ_{\eta_j}}
\def\Jjs{\cJ_{js}}
\def\Jru{\cJ_{ru}}
\def\Jrs{\cJ_{rs}}
\def\Jrr{\cJ_{\eta_r}}

\def\Gpp{{\mathcal G}_{\pi,\pi}}
\def\Gps{{\mathcal G}_{\pi,\eta_s}}
\def\Gss{{\mathcal G}_{\eta_s,\eta_s}}
\def\Gphiphi{{\mathcal G}_{\phi,\phi^\prime}}
\def\Lphiphi{\cL_{\phi,\phi^\prime}}
\def\Lpipi{\cL_{\pi,\pi}}
\def\Lssss{\cL_{\eta_s,\eta_s}}
\def\Ljjjj{\cL_{\eta_j,\eta_j}}
\def\Lrrrr{\cL_{\eta_r,\eta_r}}

\def\Jphiphi{{\cJ}_{\phi,\phi^\prime}}
\def\Jpp{{\cJ}_{\pi,\pi}}
\def\Jps{{\cJ}_{\pi,\eta_s}}
\def\Jss{{\cJ}_{\eta_s,\eta_s}}

\def\Ltphiphi{{\overline \cL}_{\phi,\phi^\prime}}
\def\Ltpipi{{\overline \cL}_{\pi,\pi}}
\def\Ltps{{\overline \cL}_{\pi,\eta_s}}
\def\Ltjjjj{{\overline \cL}_{\eta_j,\eta_j}}
\def\Ltrrrr{{\overline \cL}_{\eta_r,\eta_r}}
\def\Ltssss{{\overline \cL}_{\eta_s,\eta_s}}

%%% Local Variables: 
%%% mode: latex
%%% TeX-master: "biblio.bib"
%%% End: 

\preprint{NT@UW 04-011}

\title{Octet Baryon Masses in Partially Quenched Chiral Perturbation
  Theory}
\author{{\bf Andr\'e Walker-Loud}}
\email[]{walkloud@u.washington.edu}
\affiliation{Department of Physics, University of Washington, \\
Seattle, WA 98195-1560 }

\date{\today}

\begin{abstract} 
The mass spectrum of the lowest lying octet baryons is calculated to
next-to-next-to-leading order in heavy baryon chiral perturbation
theory and partially quenched heavy baryon chiral perturbation theory.
We work in the isospin limit and treat the
decuplet baryons as dynamical fields.  These results are necessary 
for extrapolating QCD and partially quenched QCD lattice simulations of
the octet baryon masses.

\end{abstract}

\maketitle

\section{Introduction}
  The study of low-energy QCD remains a
long-standing and exciting challenge since its conception in the early
$1970$'s.  Over the years, various models and tools have been developed
in an attempt to study the non-perturbative regime of the theory of
strong interactions.  Despite the success some of these ideas
have had, an intimate understanding of low-energy QCD and the hadron
spectra remains elusive.
Recently, progress toward this end is being made at many
fronts, one of which is lattice QCD and partially quenched chiral
perturbation theory (PQ$\chi$PT).  Lattice QCD is a first principles
method to calculate QCD observables using numerical techniques.
While ongoing progress in this area is impressive, one is presently
restricted to using light quark masses on the order of the strange
quark mass, significantly heavier than those in nature.  Hence one
needs to understand the quark mass dependence of QCD observables in 
order to compare lattice simulations with physical QCD.  For
sufficiently small quark masses, chiral perturbation theory 
($\chi$PT) can be used to extrapolate unquenched lattice simulations
from lattice quark masses to physical quark masses.  However, for
lattice masses that are presently being used the convergence of the
chiral expansion is uncertain.  Baryon masses as well as
other observables have been computed using QCD and
quenched QCD (QQCD)~%
\cite{Aoki:1999yr, AliKhan:2001tx, Allton:2001sk, Aoki:2002uc}.
In QQCD the determinant arising from the path integral over the quark
fields is set to a constant, which greatly reduces the computing time of
lattice simulations.  There has been some interest in using chiral
effective theories to extrapolate these lattice data to lighter quark
masses, for example~%
\cite{Young:2002cj, AliKhan:2003cu, Beane:2004ks}, but there is no
rigorous limit of QQCD in which one recovers QCD.

Partially quenched QCD (PQQCD) has been formulated as an
alternative to QCD and QQCD~\cite{Sharpe:1997by,
  Golterman:1998st, Sharpe:1999kj, Sharpe:2000bc}.  In PQQCD, the
valence quarks, which are coupled to external states, and the sea
quarks, which contribute to the quark determinant, are treated as 
independent fields.  Thus the valence and sea quark masses can be
different.  This freedom is generally exploited to make the valence
quark masses much smaller than the sea quark masses.  The low energy
effective theory of PQQCD is 
PQ$\chi$PT~\cite{Sharpe:2001fh, Chen:2001yi}.  The coefficients of
operators appearing in the theory are the low-energy
constants (LEC's) which characterize the short distance physics, while
the long distance behavior is characterized by meson loops.  The LEC's
of $\chi$PT are all contained in PQ$\chi$PT which allows rigorous
predictions for QCD using PQ$\chi$PT.
Moreover, the ability to independently vary the
valence and sea quark masses greatly increases the allowed parameter
space available to PQQCD.  This enables a better
extraction of the LEC's appearing in PQ$\chi$PT, as they can be fit to
a much larger set of PQ lattice calculations.

These last few years have seen much work and progress in calculating
properties of baryons in PQ$\chi$PT~%
\cite{Beane:2002vq, Chen:2002bz, Beane:2002np, Leinweber:2002qb,
  Arndt:2003vx, Beane:2003xv,Arndt:2003ww, Arndt:2003we, Arndt:2003vd}.
In this work we calculate the masses of the octet baryons to
next-to-next-to leading order (NNLO) in $\chi$PT.
We also provide the first NNLO
calculation of the octet baryon masses in PQ$\chi$PT. 
These calculations are performed in the isospin limit of $SU(3)$ and 
we keep the decuplet
baryons as dynamical fields.  In
the partially quenched calculation we use three valence, three ghost
and three sea quarks, with two of the sea quarks degenerate.
%
%%%%%%%%%  Heavy Baryon XPT  %%%%%%%%%%
%
\section{Heavy Baryon Chiral Perturbation Theory}\label{s:HBXPT}

\subsection{Pseudo-Goldstone Bosons}
  For massless quarks, QCD exhibits a chiral symmetry,
$SU(3)_L \otimes SU(3)_R \otimes U(1)_V$, which is 
spontaneously broken down to $SU(3)_V \otimes U(1)_V$.  Chiral
perturbation theory, the low-energy effective theory, emerges by
expanding about the physical vacuum state of QCD.  In the limit of
massless quarks, the pions, kaons and eta emerge as the Goldstone
bosons of the broken $SU(3)_A$.  Given that the  quark masses are
small compared to the scale of chiral symmetry breaking, the lowest
lying mesons emerge as an octet of pseudo-Goldstone bosons.

  The pseudo-Goldstone bosons are collected in an exponential matrix
\begin{equation}
  \S = 
    {\rm exp} \left( \frac{2\ i\ \Phi}{f} \right) = \xi ^2 \ ,\  
      \Phi \ =\ 
        \begin{pmatrix}
          \frac{1}{\sqrt{2}}\pi_0 + \frac{1}{\sqrt{6}}\eta & \pi^+ & K^+\\
            \pi^- & -\frac{1}{\sqrt{2}}\pi_0 + \frac{1}{\sqrt{6}}\eta & K^0\\
            K^- & \ol{K}^0 & -\frac{2}{\sqrt{6}}\eta
        \end{pmatrix}.
\label{eq:Sigma}
\end{equation}
With the above convention, the pion decay constant, $f$, is $132$~MeV.
Under an $SU(3)_L~\otimes~SU(3)_R$ transformation these fields
transform in the following way

\begin{eqnarray}
  \Sigma \rightarrow L \Sigma R^\dagger \ ,\ 
  \xi \rightarrow L \xi U^\dagger \ =\ U \xi R^\dagger ,
\label{eq:mesontrans}
\end{eqnarray}
where $U$ is implicitly defined by equations (\ref{eq:Sigma}) and
(\ref{eq:mesontrans}).  The effective Lagrangian describing the strong
dynamics of these mesons at leading order in $\chi$PT is
\begin{eqnarray}\label{eq:MessLag}
  \hL\ =\ 
    \frac{f^2}{8}{\rm Tr}\left( \partial^\mu \S ^\dagger
      \partial_\mu \S \right)
    \ +\ \l {\rm Tr} \left(m_q^\dagger\, \S ^\dagger + m_q \S \right).
\end{eqnarray}
Leading order (LO) in the power counting is $\Or (m_q)$ and thus $\Or
(m_q) \sim \Or (q^2)$, where $q$ is a pion momentum.  The quark mass
matrix, $m_q$, is given in the isospin limit, ($m_u=\md=\mbar$), by
\begin{equation}
  m_q = {\rm diag}(\mbar,\mbar,\ms).
\end{equation}
To the meson masses to LO are given by
\begin{equation}
  \mp^2 = \frac{8\l}{f^2}\,\mbar \ \ ,\ \ 
  \mk^2 = \frac{4\l}{f^2}\,(\mbar+\ms)\ \ ,\ \ 
  \me^2 = \frac{8\l}{3 f^2}\,(\mbar+2\ms)
\label{eq:mesonmass}\, .
\end{equation}
\subsection{Baryons}\label{s:HBXPTB}
  To systematically include the octet and decuplet baryons into the
chiral Lagrangian, we use heavy baryon $\chi$PT (HB$\chi$PT)~%
\cite{Jenkins:1991ne,Jenkins:1991jv,Jenkins:1991es, Jenkins:1992ts}, 
where the baryon fields are redefined in terms of velocity dependent
fields.
\begin{equation}
  B_v(x) = \frac{1+\vslash}{2} e^{i M_B \vit \cdot x} B(x)
\label{eq:Bv}\, ,
\end{equation}
where $v_\mu$ is the four-velocity of the baryon, $B$.  This field
redefinition corresponds to parameterizing the momentum of a nearly on-%
shell baryon as
\begin{equation}
  p_\mu = M_B v_\mu + k_\mu,
\label{eq:Bp}
\end{equation}
where $k_\mu$ is the residual momentum.  The effect of this is to
eliminate the standard Dirac mass term for the baryons:
\begin{equation}
  \Bbar \left(\, i\dslash - M_B \,\right) B
  = \Bbar_v \, i\dslash \, B_v 
  + \Or (\frac{1}{M_B})\, .
\label{eq:BvmB}
\end{equation}
From Eq.~(\ref{eq:Bv}), it is easy to verify that derivatives acting
on $B_v$ bring down powers of $k$, rather than $p$.  Thus, higher
dimension operators of the heavy baryon field 
$B_v$ are suppressed by powers of $M_B$.  Heavy baryon $\chi$PT is
applicable in the limit that pion momenta and the
off-shellness of the baryon are small compared to the chiral symmetry
breaking scale, $\L_\chi \sim 1 {\rm GeV}$, i.e.:
\begin{equation}
  q,\, k\cdot \vit \ll \L_\chi.
\end{equation}
In this limit, HB$\chi$PT has a consistent derivative expansion as
derivatives are suppressed by powers of $\L_\chi$.

  When the spin-$\frac{3}{2}$ decuplet baryons, $T$, are included
in the theory, an 
additional mass parameter, $\D = M_T -M_B$, must be included.  
The meson masses, $\mphi$, where $\phi$ is
a pion, kaon or eta, are much smaller than $\D$ close to the chiral limit,
and therefore the decuplet resonances can be 
integrated out of the theory, leaving only the pseudo-Goldstone
mesons and the octet baryons.  Decuplet effects would then show up
in the theory as higher dimension operators suppressed by powers of
$\C^2 / \D$ (where $\C \sim 1.5$ is the $TB\pi$ coupling).  In the
real world, $\D \sim 300$~MeV and $\mp/ \D \sim 1/2$, so
$\Or (\mp/\D)$ corrections could be sizeable and spoil the power
counting  Additionally, if one is to use present day lattice calculations to fit
the free parameters of the theory, $\mp^{latt} / \D \gtrsim 1$.
The situation becomes worse when considering $\Or (\mk/\D)$
and $\Or (\me/\D)$ corrections.  There is also ample
phenomenological evidence in the nucleon sector that suggests the
importance of including the lowest lying spin-$\frac{3}{2}$
resonances~\cite{Jenkins:1991jv,Jenkins:1991es,Jenkins:1992ts,
  Butler:1992ci,Broniowski:1993vj,Thomas:2001kw}.

  To include the decuplet baryons, one must include $\D$ in the power
counting.  The mass splitting, $\D \sim m_\phi$ so $\D$ and $m_\phi$ are $\Or
(q)$, where $q$ is a typical small pion momentum, and the quark mass, $m_q$,
is treated as $\Or (q^2)$.  The baryon mass is treated as $M_B \sim \L_\chi$.  As the
LEC's are {\it a priori} unknown, we can combine
the $1/M_B$ and $1/\L_\chi$ expansions into one expansion in powers of
$1/\L_\chi$.  There is one exception: constraints from
reparameterization invariance determine the coefficients of some of the higher
dimension operators arising in the $1/M_B$
expansion.~\cite{Luke:1992cs,Manohar:2000dt}.  Thus these $1/M_B$ 
corrections must be kept distinct to insure the Lorentz invariance of
the heavy baryon theory to a given order.

  The effective Lagrangian at leading order (LO) in the $1/M_B$
expansion  is well known~\cite{Jenkins:1991ne}.  In this paper we will
embed the baryons in rank-3 flavor tensors.  The convenience of this
choice will become readily apparent when we generalize to PQ$\chi$PT.
The $SU(3)$ matrix of the lowest-%
lying spin-$\frac{1}{2}$ baryon fields is
\begin{equation}
  {\mathbf B} = 
    \begin{pmatrix}
      \frac{1}{\sqrt{6}}\Lambda + \frac{1}{\sqrt{2}}\Sigma^0 &
        \Sigma^+ & p\\
      \Sigma^- & \frac{1}{\sqrt{6}}\Lambda -
        \frac{1}{\sqrt{2}}\Sigma^0 & n\\
      \Xi^- & \Xi^0 & -\frac{2}{\sqrt{6}}\Lambda\\
    \end{pmatrix}\, .
\label{eq:B2index}
\end{equation}
We then embed them in the tensor~\cite{Labrenz:1996jy}:
\begin{equation}
  B_{ijk}  = 
    \frac{1}{\sqrt6}\left(\ \epsilon_{ijl} {\mathbf
      B}_{k}^{\hskip 0.30em l}\ 
    +\ \epsilon_{ikl} {\mathbf B}_{j}^{\hskip 0.30em l}\ \right)\ ,
\label{eq:3indexB}
\end{equation}
which has the symmetry properties
\begin{equation}
     B_{ijk}  = B_{ikj}\ \ {\rm and}\ \ B_{ijk}+B_{jik}+B_{kji}=0.
\end{equation}
Under chiral $SU(3)$ transformations, this baryon tensor transforms as
\begin{equation}
  B_{ijk} \rightarrow U_i^{\hskip 0.30em l} U_j^{\hskip 0.30em m}
    U_k^{\hskip 0.30em n} B_{lmn}.
\end{equation}

  The spin-$\frac{3}{2}$ decuplet baryons can be described by a
Rarita-Schwinger field, $(T^\mu)_{ijk}$, which is totally symmetric
under the interchange of flavor indices, and which satisfies the
constraint, $\g_\mu~T^\mu~=~0$.  
We employ the normalization convention that $T^{111} = \Delta^{++}$.
Under a chiral $SU(3)$ transformation, the decuplet tensor transforms
identically to the octet tensor:
\begin{equation}
  T_{ijk} \rightarrow U_i^{\hskip 0.30em l} U_j^{\hskip 0.30em m}
    U_k^{\hskip 0.30em n} T_{lmn}\, .
\end{equation}  

  The octet and decuplet baryon tensors obey the spin-algebra laid out in
\cite{Jenkins:1991ne}, which can be used to eliminate the Dirac
structure of the theory; all Lorentz tensors made from spinors can be
written in terms of $v^\mu$ and $S^\mu$, where $v^\mu$ is the
four-velocity of the baryon and $S^\mu$ is the covariant spin-vector.

  Thus the Lagrangian to leading order in the $1/M_B$ expansion
can be written as \footnote{For brevity, we will drop the subscript
  $v$ from the velocity dependent heavy baryon fields.}
\begin{eqnarray}
  \hL_{LO} &=&
    \left( \Bbar\ i\vD\ B \right)\ 
    +\ 2\am \left( \Bbar B \cM_+ \right)\ 
    +\ 2\bm \left( \Bbar \cM_+ B \right)\ 
    +\ 2\smb \left( \Bbar B \right) {\rm Tr}(\cM_+)
\nonumber\\
 && -
    \left(\Tbar^{\mu}\left[\ i\vD-\Delta\right]T_\mu \right)\ 
    +\ 2\gm\left(\Tbar^\mu \cM_+ T_\mu \right)\ 
    -\ 2\smt\left(\Tbar^\mu T_\mu \right) {\rm Tr}(\cM_+)
\nonumber\\
 && +
    2\a\left(\Bbar S^\mu B A_\mu \right)\ 
    +\ 2\beta \left(\Bbar S^\mu A_\mu B \right)\ 
    +\ 2{\cal H} \left( \Tbar^\nu S^\mu A_\mu T_\nu \right)
\nonumber\\
 && +
    \sqrt{\frac{3}{2}}{\C} \left[ \left( \Tbar^\nu A_\nu B
      \right)\ 
    +\ \left( \Bbar A_\nu T^\nu \right) \right]
\label{eq:leadlag}\, .
\end{eqnarray}
Above, $D_\mu$ is the chiral-covariant derivative which acts
on the $B$ and $T$ fields as
\begin{equation}
  \left( D^\mu B \right)_{ijk} = \partial^\mu B_{ijk}\ 
    +\ (V^\mu)_i^{\hskip 0.30em l} B_{ljk}\ 
    +\ (V^\mu)_j^{\hskip 0.30em l} B_{ilk}\ 
    +\ (V^\mu)_k^{\hskip 0.30em l} B_{ijl}\, .
\end{equation}
The vector and axial-vector meson fields appearing in the Lagrangian
are given by
\begin{equation}
  V_\mu =
    \frac{1}{2} ( \xi\partial_\mu\xi^\dagger\ 
      +\ \xi^\dagger\partial_\mu\xi )\ \ ,\ \ 
  A_\mu  =
    \frac{i}{2} ( \xi\partial_\mu\xi^\dagger\ 
      -\ \xi^\dagger\partial_\mu\xi )\, ,
\end{equation}
and
\begin{equation}
  \cM_+  = \frac{1}{2}\left( \xi^\dagger m_q \xi^\dagger\ +\ \xi m_q
  \xi \right)\, .
\end{equation}
In Eq.~(\ref{eq:leadlag}) the brackets, $(~)$, denote a contraction
of the flavor and Lorentz indices.  For a matrix
$\G^\a_\beta$ acting in spin-space and a matrix $Y^i_j$ acting
in flavor space, the contractions are defined as~\cite{Labrenz:1996jy}
\begin{align}
  \left( \Bbar\ \G\ B \right) &= 
    \Bbar^{\a,kji}\ \G^\b_\a\ B_{\b,ijk}\ ,\ 
  &\left(\Tbar^\mu\ \G\ T_\mu\ \right) &= 
    \Tbar^{\mu\a,kji}\ \G^\b_\a\ T_{\mu\b,ijk}
\nonumber\\
  \left( \Bbar\ \G\ Y\ B \right) &= 
    \Bbar^{\a,kji}\ \G^\b_\a\ Y_i^{\hskip 0.30em l}\ B_{\b,ljk}\ ,
  &\left( \Tbar^\mu\ \G\ Y\ T_\mu \right) &= 
    \Tbar^{\mu\a,kji}\ \G^\b_\a\ 
      Y_i^{\hskip 0.3em l}\ T_{\mu\b,ljk}
\nonumber\\
  \left( \Bbar\ \G\ B\ Y \right) &= 
    \Bbar^{\a,kji}\ \G^\b_\a\ 
      Y_k^{\hskip 0.30em l}\ B_{\b,ijl}\ ,\ 
  &\left( \Bbar\ \G\ Y^\mu\ T_\mu \right) &= 
    \Bbar^{\a,kji}\ \G^\b_\a\ 
      \left(Y^\mu \right)_i^{\hskip 0.30em l}\
        T_{\mu\b,ljk}.
\end{align}
Such contractions ensure the proper transformations of the field
bilinears under chiral transformations.  To compare with the
coefficients used in the standard two-index baryon formulation
~\cite{Jenkins:1991ne,Jenkins:1991jv,Jenkins:1992ts}, it is
straightforward to show that:
\begin{equation}
  \a = \frac{2}{3}D + 2 F\ \ ,\ \ 
  \b = -\frac{5}{3}D +F\,, \notag\\
\end{equation}
\begin{equation}
  \am = \frac{2}{3}b_D + 2 b_F \ \ ,\ \ 
  \bm = -\frac{5}{3}b_D +b_F\ \ ,\ \ 
  \smb = b_D - b_F + \s\, .
\end{equation}
\subsection{Higher Dimensional Operators}
  The Lagrangian in Eq.~(\ref{eq:leadlag}) contains some, but not all,
terms of $\Or (q^2)$.  To calculate the baryon masses to $\Or (m_q^2)$
we must include all relevant $\Or (q^2)$ operators, which contribute to
the mass via loops, and all relevant $\Or (q^4)$ operators which
contribute at tree level.  The Lagrangian also includes operators of
$\Or (q^3)$, but they do not contribute to the baryon self-energy.

  The higher dimensional operators introduce new LEC's that, in principle,
must be determined from experiments or lattice QCD calculations.
However, many of them can be derived exactly using
reparameterisation invariance (RI) \cite{Luke:1992cs}.  The baryon
momentum parameterization in terms of $v^\mu$ and $k^\mu$ is not unique
when considering the $1/M_B$ corrections.  When the velocity and
residual momentum are simultaneously transformed in the following way
\begin{eqnarray}
  v \rightarrow v + \frac{\e}{M_B}\ \ ,\ \ 
  k \rightarrow k - \e ,
\end{eqnarray}
the momentum $p_\mu$ in Eq.~(\ref{eq:Bp}) is unchanged.
Reparameterization invariance requires the effective Lagrangian to be
invariant under such a transformation,  which ensures the
theory is Lorentz invariant to a given order in $1/M_B$.  Furthermore,
utilizing RI has 
non-trivial consequences as it connects operators at different orders
in the $1/M_B$ expansion, exactly fixing the coefficients of some of the
higher dimensional operators with respect to the lower ones.  We find
that the fixed coefficient Lagrangian is \footnote{While there are at
  least three other methods of generating the $1/M_B$ Lagrangian, the
  technique of RI allows an explicit check for redundant or missing
  operators.}
\begin{eqnarray}
  \hL^{(2)}_\frac{1}{M_B} &=& 
    -\left( \Bbar \frac{D_\perp^2}{2 M_B} B \right)\ 
      -\ \a \left( \Bbar \frac{S\cdot i \roarrow D}{M_B} B
      \vit\cdot A \right)\ 
    +\ \a \left( \Bbar \frac{i \loarrow D \cdot S}{M_B} B
      \vit\cdot A \right)
\nonumber\\
  && - \b \left( \Bbar \vit\cdot A \frac{S\cdot i \roarrow D}
       {M_B} B \right)\ 
     +\ \b \left( \Bbar \frac{i \loarrow D \cdot S}{M_B}
       \vit\cdot A B \right)
     +\ \left( \Tbar^\mu \frac{D_\perp^2}{2 M_B} T_\mu \right)
\label{eq:fixed}
\end{eqnarray}
where \footnote{Using $D^2_\perp$ instead of $D^2$ is a choice as one
  could use the LO equations of motion to eliminate the
  $(v\cdot D)^2$ term~\cite{Georgi:1991ch}, (suitably shifting the
  LECs as well so that observables remain unchanged).  This is an aesthetic
  choice as it gives the familiar form of the first relativistic
  correction to the propagator, $\vec{k}^2 /2M_B$.}
$D_\perp^2 = D^2 - (\vit \cdot D)^2$.

The Lagrangian contains additional $\Or (q^2)$ operators as well
as $\Or (q^4)$ operators which are invariant under the $SU(3)$ chiral
transformations.  The operators relevant to the self energy of the
octet baryons are
\begin{eqnarray}
     \hL^{(2,4)}_{c.t.} &=& \frac{1}{( 4\pi f)}\Bigg\{
          \ \ b^A_1\, \Bbar^{kji} \left( A\cdot A \right)_i^{\hskip0.3em n} 
               B_{njk}\ 
          +\ b^A_2\, \Bbar^{kji} \left( A\cdot A \right)_k^{\hskip0.3em
               n} B_{ijn} \nonumber\\
       && \phantom{4\pi f an}
          \ + b^A_3\, \Bbar^{kji} (A_\mu )_i^{\hskip0.3em l}
               (A^\mu )_j^{\hskip0.3em n} B_{lnk}\ 
          +\ b^A_4\, \Bbar^{kji} B_{ijk} {\rm Tr}\left( A\cdot A
          \right) \nonumber\\
       && \phantom{4\pi f an}
          \ + b^{vA}_1\, \Bbar^{kji} \left( v\cdot\, A\, v\cdot A
           \right)_i^{\hskip0.3em n} B_{njk}\ 
          +\ b^{vA}_2\, \Bbar^{kji} \left( v\cdot A\, v\cdot A
               \right)_k^{\hskip0.3emn} B_{ijn} \nonumber\\ 
       && \phantom{4\pi f an}
          \ + b^{vA}_3\, \Bbar^{kji} (v\cdot A )_i^{\hskip0.3em l}
               (v\cdot A )_j^{\hskip0.3em n} B_{lnk}\ 
          +\ b^{vA}_4\, \Bbar^{kji} B_{ijk} {\rm Tr}\left( v\cdot A\, v\cdot A
          \right) \nonumber\\
       && \phantom{4\pi f an}
          \ + b^M_1\, \Bbar^{kji} \left( \cM_+ \cM_+ \right)_i^{\hskip0.3em n} 
               B_{njk}\ 
          +\ b^M_2\, \Bbar^{kji} \left( \cM_+ \cM_+ \right)_k^{\hskip0.3em n} 
               B_{ijn}
\nonumber\\
       && \phantom{4\pi f an}
          \ + b^M_3\, \Bbar^{kji} (\cM_+)_i^{\hskip0.3em l}
               (\cM_+)_j^{\hskip0.3em n} B_{lnk}\ 
          +\ b^M_4\, \Bbar^{kji} B_{ijk} {\rm Tr}\left(\cM_+ \cM_+\right)\ 
\nonumber\\
       && \phantom{4\pi f an}
          \ + b^M_5\, \Bbar^{kji} (\cM_+)_i^{\hskip0.3em n} B_{njk} 
               {\rm Tr}\left( \cM_+ \right)\ 
          +\ b^M_6\, \Bbar^{kji} (\cM_+)_k^{\hskip0.3em n} B_{ijn} 
               {\rm Tr}\left( \cM_+ \right)
\nonumber\\
       && \phantom{4\pi f an}
          \ + b^M_7\, \Bbar^{kji} B_{ijk} {\rm Tr}\left( \cM_+ \right)
               {\rm Tr}\left( \cM_+ \right) \Bigg\}
\label{eq:hdo}\, .
\end{eqnarray}

The operators of the form $\big( \Bbar v\cdot A v\cdot A B \big)$ have
not been kept explicit in the literature so far for calculations of
the octet baryon masses to $\Or (m_q^2)$~\cite{Jenkins:1991jv,
  Jenkins:1992ts,Borasoy:1997bx}.
These operators have identical flavor structure to the corresponding
$\big( \Bbar A\cdot A B \big)$ operators.  However, these two sets of
operators have different Lorentz structure, which gives rise to
different finite $\mphi^4$ contributions to the octet baryon self
energy calculations.  One can choose a renormalization scheme such
that the contributions to the octet baryon masses from these different
operators can not be distinguished.  Therefore, with a suitable
redefinition of the $b^A_{1-4}$ and $b^M_{1-7}$ coefficients, the
operators with coefficients $b^{vA}_{1-4}$ can be neglected in the
baryon mass calculations, as their contributions to the masses can be
absorbed by the other operators to this order in the chiral expansion.

However, the operators with coefficients $b^A_{1-4}$ and
$b^{vA}_{1-4}$ can be distinguished in $\pi N \rightarrow \pi N$
scattering at tree level, for example. It is therefore useful to keep both
types of operators explicitly in the Lagrangian, allowing them to be
distinguished in the octet baryon mass calculation.
This also provides a consistent means to determine these LECs, as they
are the same coefficients which appear in the Lagrangian used for the
calculation of other observables like $\pi N \rightarrow \pi N$
scattering.~\footnote{One could also add operators of the form
  $\big( \Bbar S\cdot A S\cdot A B \big)$, but it is straight forward
  to show that they are a linear combination of the operators in
  Eq.~(\ref{eq:hdo}), and therefore not distinct.}

In principle, additional $1/M_B$ operators with the same chiral
symmetry properties as those contained in Eq.~(\ref{eq:hdo}) can be
generated.  However, these $1/M_B$ operators do not have their
coefficients constrained by RI, therefore they can be absorbed by a
re-definition of the $b^{A,vA,M}_i$ coefficients.  For example, the
Rarita-Schwinger field used to describe the decuplet baryons contains
un-physical spin-$1/2$ degrees of freedom which can propagate when the
decuplet baryons are off mass-shell.  It can be shown that the
contribution to the octet baryon masses from these un-physical degrees
of freedom is suppressed by $1/M_B$~\cite{Hemmert:1997ye}.
Therefore these effects are implicitly included in the operators with
coefficients $b^{A,vA}_{1-4}$.

The full Lagrangian relevant to the calculation of the octet baryon
masses to $\Or (m_q^2)$ is
\begin{equation}
  \hL_{m_q^2} = \hL_{LO} + \hL^{(2)}_\frac{1}{M_B} + \hL^{(2,4)}_{c.t.}\,
.\end{equation}
\section{Baryon Masses}\label{s:XPTmasses}
The mass of the $i^{th}$ baryon in the chiral expansion is
\begin{eqnarray}
     M_{B_i} = M_0\left(\mu \right) \ -\ M_{B_i}^{(1)}\left(\mu \right)
               \ -\ M_{B_i}^{(3/2)}\left(\mu \right)
               \ -\ M_{B_i}^{(2)}\left(\mu \right)\ +\ .\, .\ .
\label{eq:massexp}
\end{eqnarray}
  Here, $M_0 \left(\mu \right)$ is the renormalized mass of the octet
baryons in the chiral limit, is independent of $m_q$ and also $B_i$.
$M_{B_i}^{(n)}$ is the contribution to the 
$i^{th}$ octet baryon of the order $m_q^{(n)}$, and $\mu$ is the
dimensional regularization scale.  For this calculation we use a
renormalization scheme of always subtracting terms proportional to
\begin{equation}
\bigg[ \frac{1}{\e} -\g +1 +\ln 4\pi \bigg]. \notag
\end{equation}

The inclusion of the decuplet baryon fields in HB$\chi$PT requires
additional operators involving powers of $\D / \L_\chi$, as $\D$ is a
chiral singlet.  Thus all terms in the Lagrangian can be multiplied by
arbitrary functions of $\D / \L_\chi$ without altering their symmetry
properties under chiral transformations.  To the order we are
working, we can systematically include the contributions to the octet
masses from these additional operators, by treating all constants in
the calculation as arbitrary polynomial functions of $\D / \L_\chi$,
and expand them to the appropriate order.  For example $\am, \bm$ and
$\smb$ must be expanded out to $\Or (\D^2/\L^2_\chi)$ while $D,F$ and $\C$
must be expanded to $\Or (\D/\L_\chi)$,
\begin{eqnarray}
  \am \rightarrow \am\left(\frac{\D}{\L_\chi}\right) &=& 
      \am\left[1+a^M_1 \left(\frac{\D}{\L_\chi} \right)
      + a^M_2 \left(\frac{\D}{\L_\chi}\right)^2 
      +\Or\left(\frac{\D^3}{\L_\chi^3}\right)
      \right], \nonumber\\
  D \rightarrow D \bigg( \frac{\D}{\L_\chi} \bigg) &=&
      D \left[ 1 + d_1 \bigg( \frac{\D}{\L_\chi} \bigg) 
      + \Or \bigg( \frac{\D^2}{\L_\chi^2} \bigg) \right].
\end{eqnarray}
Additionally, the LECs are also implicit functions of $\mu$, to
cancel the scale dependence of observables arising from the loop integrals.

\begin{figure}[tb]
  \centering
  \includegraphics[width=0.50\textwidth]{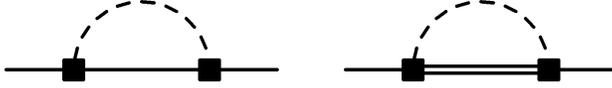}%
  \caption{
    One loop graphs which give contributions to $M_{B_i}^{(3/2)}$.  The
    single, double and dashed lines correspond to octet baryons,
    decuplet baryons and mesons, respectively.  The filled squares
    denote the axial coupling given in Eq.~(\ref{eq:leadlag}).
  }
  \label{fig:NLO}
\end{figure}

\begin{figure}[tb]
  \centering
  \includegraphics[width=0.75\textwidth]{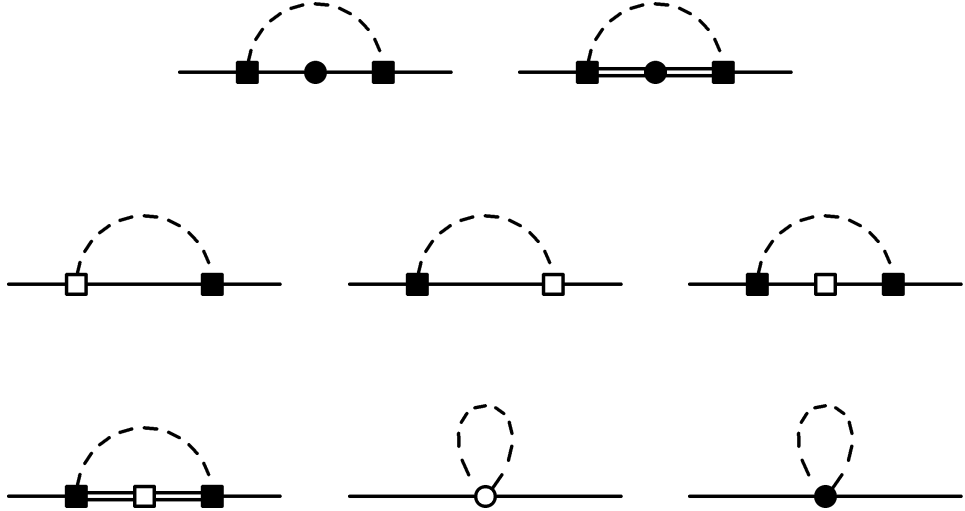}%
  \caption{
    One-loop graphs which give contributions to $M_{B_i}^{(2)}$.   
    Single, double and dashed lines correspond to octet baryons,
    decuplet baryons and mesons, respectively.  The filled squares and
    filled circles denote axial couplings and insertions of $\cM_+$ given
    in Eq.~(\ref{eq:leadlag}).  The empty squares denote insertions of
    fixed $1/M_B$ operators from Eq.~(\ref{eq:fixed}), and the empty
    circles denote two-baryon-axial couplings defined in
    Eq.~(\ref{eq:hdo}).
  }
  \label{fig:NNLO}
\end{figure}

  In calculating the masses, the leading dependence upon $m_q$ arises
from the terms in our 
Lagrangian, Eq. (\ref{eq:leadlag}), with coefficients $\am, \bm, {\rm
  and}\ \smb$, which can be determined from experiment or lattice
simulations.  The 
$\Or (m_q^{3/2})$ contributions arise from the one-loop diagrams shown
in Fig.~\ref{fig:NLO},
which are formed from the operators in the Lagrangian with coefficients
$\a, \b$ and $\C$.  The $\Or (m_q^2)$ contributions arise from the
one-loop diagrams shown in 
Fig.~\ref{fig:NNLO},
from the tree-level contributions of the operators with coefficients,
$b^M_i$, and from the NLO wave-function corrections.

  We find that the contributions to the nucleon mass are \footnote{We
  express $\a$ and $\b$ in terms of $D$ and $F$, following  the
  convention set by \cite{Chen:2001yi}.}
%
%%%%%%%%%%%%%%  Nucleon Mass %%%%%%%%%%%%%%%%%%
%
\begin{eqnarray}\label{eq:NmassLO}
  M_N^{(1)} &=&
       2\mbar ( \am + \bm + 2\smb )\ +\ 2 m_s \smb\, ,\\
\label{eq:NmassNLO}
  M_N^{(3/2)} &=&
       \frac{1}{8\pi f^2}\bigg[
         \frac{3}{2}\left(D+F\right)^2 \mp^3\ 
         +\ \frac{1}{6} \left( D-3 F \right)^2 \me^3\ 
         +\ \frac{1}{3} \left(5D^2-6DF+9F^2\right) \mk^3
\nonumber\\
  &&\qquad\quad
       \ +\ \frac{\C^2}{\pi}\left( \frac{4}{3}\cF (\mp,\D,\mu ) +
         \frac{1}{3} \cF (\mk,\D,\mu) \right) \bigg],
\end{eqnarray}
consistent with previous calculations~\cite{Jenkins:1991ne,
  Bernard:1993nj}.  The contribution of this work is
the NNLO mass contribution to the octet baryons in $SU(3)$,
\begin{eqnarray}
  -M_N^{(2)} &=&
   -\overline \cL (\mp,\mu) \bigg\{
     \frac{1}{(4\pi f)^3}\bigg[
      \frac{3}{2}b^A_1 + \frac{3}{2}b^A_2 - b^A_3 +3b^A_4 
      +\frac{3}{8}b^{vA}_1 + \frac{3}{8}b^{vA}_2 -\frac{1}{4}b^{vA}_3
       +\frac{3}{4}b^{vA}_4 \bigg] \nonumber\\
  &&\qquad\qquad\qquad
     +\frac{1}{M_B(4\pi f)^2} \bigg[
      \frac{27}{16}(D+F)^2 +\frac{5}{2}C^2 \bigg] \bigg\} \nonumber\\
  &&-\overline \cL (\me,\mu) \bigg\{
     \frac{1}{(4\pi f)^3} \bigg[
      \frac{1}{6}b^A_1 +\frac{1}{6}b^A_2 +\frac{1}{6}b^A_3 +b^A_4 
       +\frac{1}{24}b^{vA}_1 +\frac{1}{24}b^{vA}_2
       +\frac{1}{24}b^{vA}_3 +\frac{1}{4}b^{vA}_4 \bigg] \nonumber\\
  &&\qquad\qquad\qquad
     +\frac{1}{M_B(4\pi f)^2} \frac{3}{16}(D-3F)^2 \bigg\} \nonumber\\
  &&-\overline \cL (\mk,\mu) \bigg\{
      \frac{1}{(4\pi f)^3} \bigg[
       b^A_1 + b^A_2 + 4b^A_4 +
       \frac{1}{4}b^{vA}_1 +\frac{1}{4}b^{vA}_2 +b^{vA}_4 \bigg]
       \nonumber\\
  &&\qquad\qquad\qquad
     +\frac{1}{M_B(4\pi f)^2} \bigg[
       \frac{3}{8}(5D^2-6DF+9F^2) +\frac{5}{8}C^2 \bigg] \bigg\}
\nonumber\\
  &&+\cL (\mp,\mu)\ \frac{\mbar\ 6(\am +\bm + 2\smb )}{(4\pi f)^2}
     +\cL (\me,\mu) \frac{1}{(4\pi f)^2} \bigg[ \frac{2}{3}\mbar (\am
     +\bm +2\smb) +\frac{8}{3}\ms \bigg] \nonumber\\
  &&+\cL (\mk,\mu) \frac{1}{(4\pi f)^2}\bigg\{
       2(\mbar +\ms)(\am +\bm +4\smb) +(5D^2-6DF+9F^2) \MN \nonumber\\
  &&\qquad\qquad\qquad\qquad\qquad
        -\frac{9}{2}(D-F)^2 \MS -\frac{1}{2}(D+3F)^2 \ML \bigg\}
        \nonumber\\
  && +\cJ (\mp,\D,\mu) \frac{4\C^2}{(4\pi f)^2} \left[\MN +\MD \right]
     +\cJ (\mk,\D,\mu) \frac{\C^2}{(4\pi f)^2} \left[\MN +\MSS \right] \nonumber\\
  && +\mp^4 \bigg\{
       \frac{1}{(4\pi f)^3} \bigg[ \frac{3}{16}b^{vA}_1 +
         \frac{3}{16}b^{vA}_2 -\frac{1}{8}b^{vA}_3 +\frac{3}{8}b^{vA}_4
         \bigg]
       +\frac{1}{M_B(4\pi f)^2} \bigg[ \frac{45}{32}(D+F)^2 
         +\frac{9}{4}C^2 \bigg] \bigg\} \nonumber\\
  && +\mp^2 \frac{4C^2}{(4\pi f)^2} \bigg[ \MN +\MD \bigg] \nonumber\\
  && +\me^4 \bigg\{
       \frac{1}{(4\pi f)^3} \bigg[ \frac{1}{48}b^{vA}_1
         +\frac{1}{48}b^{vA}_2 +\frac{1}{48}b^{vA}_3
         +\frac{1}{8}b^{vA}_4 \bigg]
       +\frac{1}{M_B(4\pi f)^2} \bigg[ \frac{5}{32}(D-3F)^2 \bigg]
       \bigg\} \nonumber\\
  && +\mk^4 \bigg\{
       \frac{1}{(4\pi f)^3} \bigg[ \frac{1}{8}b^{vA}_1
         +\frac{1}{8}b^{vA}_2 +\frac{1}{2}b^{vA}_4 \bigg]
       +\frac{1}{M_B(4\pi f)^2} \bigg[ \frac{5}{16}(5D^2-6DF+9F^2)
         +\frac{9}{16} C^2 \bigg] \bigg\} \nonumber\\
  && +\mk^2 \frac{1}{(4\pi f)^2} \bigg\{
       \frac{2}{3}(5D^2-6DF+9F^2)\MN -3(D-F)^2\MS \nonumber\\
  &&\qquad\qquad\qquad\quad
      -\frac{1}{3}(D+3F)^2\ML +C^2 \Big[ \MN +\MSS \Big] \bigg\}\nonumber\\
  && +\frac{\mbar^2}{(4\pi f)} \Big[
       b^M_1 +b^M_2 +b^M_3 +2b^M_4 +2b^M_5 +2b^M_6 +4b^M_7 \Big]
       \nonumber\\
  && +\frac{\mbar \ms}{(4\pi f)} \Big[
       b^M_5 +b^M_6 +4b^M_7 \Big]
     +\frac{\ms^2}{(4\pi f)} \Big[ b^M_4 +b^M_7 \Big]
\label{eq:NmassNNLO}
\end{eqnarray}
The LO octet and decuplet masses
given in the above expression can be found in
Appendix~\ref{ap:XPTmasses}.  We have made use of
several abbreviations for the functions entering from
loop contributions, namely:
\begin{eqnarray}\label{eq:Xlogs1}
  \cL (\mphi,\mu) &=& \mphi^2\ \ln \left( \frac{\mphi^2}{\mu^2} \right),
    \nonumber\\
  \overline \cL (\mphi,\mu) &=& \mphi^4 \ \ln \left(
    \frac{\mphi^2}{\mu^2} \right), \nonumber\\
  \cF (\mphi,\D,\mu) &=& 
    \left( \mphi^2 - \D^2 \right) 
      \left[ \sqrt{\D^2 - \mphi^2}\ 
        \ln \left( \frac{\D - \sqrt{\D^2 - \mphi^2 + i\e}} 
                        {\D + \sqrt{\D^2 - \mphi^2 + i\e}} \right) - 
      \D \ln \left(\frac{\mphi^2}{\mu^2}\right) \right]\nonumber\\
    && \qquad\qquad\qquad\qquad
    - \frac{1}{2}\D \mphi^2 \ln \left(\frac{\mphi^2}{\mu^2}
      \right)\, ,\nonumber\\
  \cJ \left(\mphi,\D,\mu \right) &=&
    \left( \mphi^2 - 2\D^2 \right) \ln
    \left(\frac{\mphi^2}{\mu^2}\right)
\nonumber\\
    && \qquad\qquad
      +2\D \sqrt{\D^2 -
      \mphi^2}\ \ln \left( \frac{\D - \sqrt{\D^2 - \mphi^2 +
          i\e}} {\D + \sqrt{\D^2 - \mphi^2 + i\e}} \right)\, .
\end{eqnarray}

It should be stressed that the above expansion,
Eq.~(\ref{eq:NmassLO}), (\ref{eq:NmassNLO}) and (\ref{eq:NmassNNLO}),
is a quark mass expansion.  The meson masses are replacements for the
quark masses, given by Eq.~(\ref{eq:mesonmass}).  If an expansion in
terms of the physical meson masses is desired, then to be consistent
to $\Or(m_q^2) \sim \Or(\mphi^4)$, one needs to use the NLO expression for
the meson masses in Eq.~(\ref{eq:NmassLO}) and the LO expression in
Eq.~(\ref{eq:NmassNLO}) and Eq.~(\ref{eq:NmassNNLO}).  These results
build on previous work~\cite{Jenkins:1992ts, Lebed:1994gt, Banerjee:1995bk, Borasoy:1997bx}.
The results for the $\chi$PT calculation of the $\S, \L$ and
$\X$ baryons are located in Appendix~\ref{ap:XPTmasses}.
%
%
%
%%%%%%%%%  The PQXPT L's  %%%%%%%%%%%%
%
\section{PQ$\chi$PT}
In PQQCD Lagrangian is
\begin{equation}
\hL = \sum_{j,k=1}^9 \ol{Q}^{\hskip 0.2em j} \left[
  i\Dslash - m_Q \right]_j^{\hskip 0.3em k} Q_k.
\label{eq:pqqcdlag}
\end{equation}
This differs from the $SU(3)$ Lagrangian of QCD by the
inclusion of six extra quarks; three bosonic ghost quarks, ($\tilde u,
\tilde d, \tilde s$), and three fermionic sea quarks, ($j, l, r$).  The
nine quark fields transform in the fundamental representation of the
graded $SU(6|3)$ group.  They are combined in the nine-component vector
\begin{equation}
  Q=
  \begin{pmatrix}
    q\\ q_{sea}\\ \tilde q\\
  \end{pmatrix}.
\end{equation}
Here we have separated the quark field vector into the valence, sea
and ghost sectors:
\begin{equation}
  q=
  \begin{pmatrix}
    u\\ d\\ s\\
  \end{pmatrix} ,\ 
  q_{sea}=
  \begin{pmatrix}
    j\\ l\\ r\\
  \end{pmatrix} ,\ 
  {\tilde q}=
  \begin{pmatrix}
    \tilde u\\ \tilde d\\ \tilde s\\
  \end{pmatrix}.
\end{equation}
The quark fields obey the equal-time commutation relation
\begin{equation}
Q^\a_i(\mathbf x) Q^{\beta \dagger}_j(\mathbf y) -
(-1)^{\eta_i \eta_j} Q^{\b \dagger}_j(\mathbf y) Q^\a_i(\mathbf x) =
\d^{\a \b} \d_{ij} \d^3 (\mathbf {x-y}),
\end{equation}
where $\a, \beta$ are spin and $i,j$ are flavor indices.
Analogous graded equal-time commutation relations can be written for
two $Q$'s and two $Q^\dagger$'s.  The grading factors,
\begin{equation}
   \eta_k
   = \left\{ 
       \begin{array}{cl}
         1 & \text{for } k=1,2,3,4,5,6 \\
         0 & \text{for } k=7,8,9
       \end{array}\, ,
     \right.
\end{equation}
take into account the different fermionic and bosonic statistics for
the quark fields.  In the isospin limit the quark mass matrix of
$SU(6|3)$ is given by
\begin{equation}
  m_Q = {\rm diag}(\mbar, \mbar, m_s, m_j, m_j, m_r, \mbar, \mbar, m_s).
\end{equation}
Setting the ghost quark masses equal to the valence quark masses
results in an exact cancellation in the path integral between the
valence quark determinant and the ghost quark determinant.   The sea
quark determinant is unaffected. Thus one has a way to vary the
valence and sea quark masses independently.  
%This feature is powerful for modern lattice simulations.  
QCD is recovered in the
limit $m_j \rightarrow \mbar$ and $m_r \rightarrow m_s$.
\subsection{Pseudo-Goldstone Mesons}
For massless quarks, the theory corresponding to the Lagrangian in
Eq. (\ref{eq:pqqcdlag}) has a 
graded $SU(6|3)_L \otimes SU(6|3)_R \otimes U(1)_V$ symmetry which is
assumed to be spontaneously broken down to $SU(6|3)_V \otimes U(1)_V$
in analogy with QCD.  The effective low-energy theory obtained by
perturbing about the physical vacuum state of PQQCD is PQ$\chi$PT.
The result is 80~pseudo-Goldstone mesons whose dynamics can be
described at LO in the chiral expansion by the Lagrangian
\begin{equation}
  \hL =
    \frac{f^2}{8} {\rm str} \left(
      \partial^\mu \S^\dagger \partial_\mu \S \right)
      + \l {\rm str} \left( m_Q \S + m_Q^\dagger \S^\dagger \right)
           +\a_\Phi \partial^\mu \Phi_0 \partial_\mu \Phi_0
           - m_0^2 \Phi_0^2,
\label{eq:pqbosons}
\end{equation}
where
\begin{equation}
  \S =
    {\rm exp} \left( \frac{2 i \Phi}{f} \right) = \xi^2,\ 
    \Phi =
    \begin{pmatrix}
      M & \chi^\dagger\\
      \chi & \tilde M\\
    \end{pmatrix}.
\end{equation}
The operation str(\ ) in Eq. (\ref{eq:pqbosons}) is the
supertrace over flavor indices.  The quantities $\a_\Phi$ and
$m_0$ are non-vanishing in the chiral limit.  $M$ and $\tilde M$ are
matrices containing bosonic mesons while $\chi$ and $\chi^\dagger$
are matrices containing fermionic mesons:
\begin{eqnarray}
M &=&
      \begin{pmatrix}
      \eta_u & \pi^+ & K^+ & J^0 & L^+ & R^+\\
      \pi^- & \eta_d & K^0 & J^- & L^0 & R^0\\
      K^- & {\ol K}^0 & \eta_s & J^-_s & L^0_s & R^0_s\\
      {\ol J}^0 & J^+ & J^+_s & \eta_j & Y^+_{jl} & Y^+_{jr}\\
      L^- & {\ol L}^0 & {\ol L}^0_s & Y^-_{jl} & \eta_l &
               Y^0_{lr}\\
      R^- & {\ol R}^0 & {\ol R}^0_s & Y^-_{jr} &
               {\ol Y}^0_{lr} & \eta_r\\
      \end{pmatrix}\ ,\ 
\tilde M =
      \begin{pmatrix}
      {\tilde \eta}_u & {\tilde \pi}^+ & {\tilde K}^+\\
      {\tilde \pi}^- & {\tilde \eta}_d & {\tilde K}^0\\
      {\tilde K}^- & {\tilde {\ol K}}^0 & {\tilde \eta}_s\\
      \end{pmatrix}
\nonumber\\
\chi &=&
      \begin{pmatrix}
      \chi_{\eta_u} & \chi_{\pi^+} & \chi_{K^+} & \chi_{J^0} &
               \chi_{L^+} & \chi_{R^+}\\
      \chi_{\pi^-} & \chi_{\eta_d} & \chi_{K^0} & \chi_{J^-} &
               \chi_{L^0} & \chi_{R^0}\\
      \chi_{K^-} & \chi_{{\ol K}^0} & \chi_{\eta_s} &
               \chi_{J^-_s} & \chi_{L^0_s} & \chi_{R^0_s}\\
      \end{pmatrix}.
\end{eqnarray}
The upper $3 \times 3$ block of $M$ is the usual octet of
pseudo-scalar mesons and the remaining components are mesons formed
with one or two sea quarks.

  The flavor singlet field is defined to be $\Phi_0 = {\rm str}( \Phi ) /
{\sqrt 6}$.  PQQCD has a strong axial anomaly, $U(1)_A$,
and therefore the mass of the singlet field, $m_0$ can be taken to be
the order of the chiral symmetry breaking scale, $m_0 \rightarrow
\Lambda_\chi$\cite{Sharpe:2000bn}.  In this limit, the $\eta$
two-point correlation functions deviate from the simple, single pole
form.  The $\eta_a \eta_b$ propagator with $2+1$ sea-quarks and $a,b =
u,d,s,j,l,r,\tilde u, \tilde d, \tilde s$ is found to be at leading
order:
\begin{equation}
{\cal G}_{\eta_a \eta_b} =
        \frac{i \epsilon_a \delta_{ab}}{q^2 - m^2_{\eta_a} +i\epsilon}
        - \frac{i}{3} \frac{\epsilon_a \epsilon_b \left(q^2 - m^2_{jj}
            \right) \left( q^2 - m^2_{rr} \right)}
            {\left(q^2 - m^2_{\eta_a} +i\epsilon \right)
             \left(q^2 - m^2_{\eta_b} +i\epsilon \right)
             \left(q^2 - m^2_X +i\epsilon \right)}\, ,
\end{equation}
where
\begin{equation}
\epsilon_a = (-1)^{1+\eta_a}
\end{equation}
The mass, $m_{xy}$, is the mass of a meson composed of (anti)-quarks
of flavor $x$ and $y$, while the mass, $m_X$ is defined as $m_X^2 =
\frac{1}{3}\left(m^2_{jj} + 2   m^2_{rr}\right)$.  This can be
compactly written as
\begin{equation}
{\cal G}_{\eta_a \eta_b} =
         \e_a \d_{ab} P_a +
         \e_a \e_b {\cal H}_{ab}\left(P_a,P_b,P_X\right),
\end{equation}
where
\begin{eqnarray}
     P_a &=& \frac{i}{q^2 - m^2_{\eta_a} +i\e}\ ,\ 
     P_b = \frac{i}{q^2 - m^2_{\eta_b} +i\e}\ ,\ 
     P_X = \frac{i}{q^2 - m^2_X +i\e}
\nonumber\\
\nonumber\\
\nonumber\\
     {\cal H}_{ab}\left(A,B,C\right) &=& 
           -\frac{1}{3}\left[
             \frac{\left( m^2_{jj}-m^2_{\eta_a}\right)
                   \left( m^2_{rr}-m^2_{\eta_a}\right)}
                  {\left( m^2_{\eta_a} - m^2_{\eta_b}\right)
                   \left( m^2_{\eta_a} - m^2_X\right)}
                 A
            -\frac{\left( m^2_{jj}-m^2_{\eta_b}\right)
                   \left( m^2_{rr}-m^2_{\eta_b}\right)}
                  {\left( m^2_{\eta_a} - m^2_{\eta_b}\right)
                   \left( m^2_{\eta_b} - m^2_X\right)}
                 B \right.\, ,
\nonumber\\
&&\qquad\quad\left.
            +\frac{\left( m^2_X-m^2_{jj}\right)
                   \left( m^2_X-m^2_{rr}\right)}
                  {\left( m^2_X-m^2_{\eta_a}\right)
                   \left( m^2_X-m^2_{\eta_b}\right)}
                 C\ \right].
\label{eq:Hfunction}
\end{eqnarray}

\subsection{Baryons}
In PQ$\chi$PT the baryons are composed of three quarks, $Q_iQ_jQ_k$,
where $i-k$ can be valence, sea or ghost quarks. 
One decomposes the irreducible representations of $SU(6|3)_V$ into
irreducible representations of $SU(3)_{val} \otimes SU(3)_{sea}
\otimes SU(3)_{ghost} \otimes U(1)$.  The method for including the
octet and decuplet baryons into PQ$\chi$PT is to use the interpolating
field~\cite{Labrenz:1996jy,Chen:2001yi}:
\begin{equation}
  \cB_{ijk}^\g \sim
    \left(Q_i^{\a,a} Q_j^{\beta,b} Q_k^{\g,c}-Q_i^{\a,a}
    Q_j^{\g,c} Q_k^{\beta,b}\right)
    \epsilon_{abc}(C\g_5)_{\a \beta}.
\end{equation}
We require that $\cB_{ijk} = B_{ijk}$, defined in
Eq.~(\ref{eq:3indexB}), when the indices, $i,j,k$ are restricted to
$1-3$.  Thus the octet baryons are contained as an $(\mathbf{8,1,1})$ 
of $SU(3)_{val} \otimes SU(3)_{sea} \otimes SU(3)_{ghost} \otimes
U(1)$ in the $\mathbf{240}$ representation.  In addition to the
conventional octet baryons composed of valence quarks, $\cB_{ijk}$
also contains baryon fields composed of sea and ghost quarks.  In this
paper we only need the baryons which contain at most one
sea or ghost quark, and these states have been explicitly constructed
in \cite{Chen:2001yi}.  Under the interchange of flavor indices, one
finds~\cite{Labrenz:1996jy}:
\begin{equation}
\cB_{ijk} = (-)^{1+\eta_j \eta_k}\cB_{ikj}
           \quad {\rm and} \quad
  \cB_{ijk} + (-)^{1+\eta_i \eta_j}\cB_{jik}
                 + (-)^{1+\eta_i \eta_j + \eta_j \eta_k + \eta_i
                   \eta_k} \cB_{kji}
                 =0.
\end{equation}
Similarly, one can construct the spin-$\frac{3}{2}$ decuplet baryons
which are embedded in the $\mathbf{138}$, and have an interpolating
field
\begin{equation}
  \cT_{ijk}^{\a,\mu} \sim
      \left( Q_i^{\a,a} Q_j^{\beta,b} Q_k^{\g,c}
            +Q_i^{\beta,b} Q_j^{\g,c} Q_k^{\a,a}
            +Q_i^{\g,c} Q_j^{\a ,a} Q_k^{\beta,b}
      \right)
         \epsilon_{abc} \left(C\g^\mu \right)_{\beta \g}.
\end{equation}
We require that $\cT_{ijk} = T_{ijk}$, when the indices $i,j,k$ are
restricted to $1-3$.  Under $SU(3)_{val} \otimes SU(3)_{sea} \otimes
SU(3)_{ghost} \otimes U(1)$ they transform as a $(\mathbf{ 10,1,1})$.
In addition to the conventional decuplet resonances composed of
valence quarks, $\cT_{ijk}$ contains fields with sea and ghost
quarks, which have also been constructed in \cite{Chen:2001yi}.

Under the interchange of flavor indices, one finds that
\begin{equation}
\cT_{ijk} = (-)^{1 + \eta_i \eta_j}\cT_{jik} = 
                 (-)^{1 + \eta_j \eta_k}\cT_{ikj}\, .
\end{equation}
Under $SU(6|3)_V$, both $\cB_{ijk}$ and ${\cal
  T}_{ijk}$ transform as \cite{Labrenz:1996jy}
\begin{equation}
\cB_{ijk} \rightarrow
       (-)^{\eta_l (\eta_j +\eta_m)+(\eta_l + \eta_m)(\eta_k +
         \eta_n)} U_i^{\hskip 0.3em l} U_j^{\hskip 0.3em m} 
                  U_k^{\hskip 0.3em n}
               \cB_{lmn}\, .
\end{equation}

\subsection{Baryon-Meson Lagrange Density}
The utility of using 3-index baryon tensors in the QCD
calculation is fully realized when one extends the $\chi$PT Lagrangian to
PQ$\chi$PT.  It has recently been shown that in the meson sector there
are extra operators in PQ$\chi$PT, that do
not arise in $\chi$PT~\cite{Sharpe:2003vy}.  The
Cayley-Hamilton identity for traceless $3 \times 3$ matrices, allows
one to reduce the number of operators in the meson sector of $\chi$PT.
There is no such identity for the supertraceless matrices of $SU(6|3)$,
allowing for extra operators.  Use of the $SU(3)$ matrix of
baryon fields, Eq.~(\ref{eq:B2index}), raises similar concerns at
higher orders in the expansion of the Lagrangian, when extending to
PQ$\chi$PT.  However, when using three-index baryon tensors, the
flavor contractions involving the baryon fields can not be expressed
as a trace, as with the more familiar form~\cite{Jenkins:1991ne,
  Jenkins:1991jv, Jenkins:1991es, Jenkins:1992ts, Borasoy:1997bx,
  Bernard:1993nj}.  Therefore, there is no danger of inadvertently 
neglecting operators when extending from $SU(3)$ HB$\chi$PT to $SU(6|3)$
PQHB$\chi$PT.  Furthermore, 
when employing the three-index notation for the baryon fields the
coefficients of the HB$\chi$PT operators are the same
as the coefficients of the corresponding PQHB$\chi$PT
operators~\cite{Sharpe:2000bn}.  This is useful as for the 
foreseeable future, reliable calculations of baryonic matrix elements
will be performed with PQQCD lattice simulations.

  To write down the PQ$\chi$PT Lagrangian, we must also include the
appropriate grading factors.  The flavor contractions are now defined
as~\cite{Chen:2001yi}
\begin{align}
  \left( \cBbar\ \G\ \cB \right) &= 
    \cBbar^{\a,kji}\ \G^\b_\a\ \cB_{\b,ijk}\ ,\ 
  &\left(\cTbar^\mu\ \G\ \cT_\mu\ \right) &= 
    \cTbar^{\mu\a,kji}\ \G^\b_\a\ \cT_{\mu\b,ijk}
\nonumber\\
  \left( \cBbar\ \G\ Y\ \cB \right) &= 
    \cBbar^{\a,kji}\ \G^\b_\a\ Y_i^{\hskip 0.30em l}\ \cB_{\b,ljk}\ ,
  &\left( \cTbar^\mu\ \G\ Y\ \cT_\mu \right) &= 
    \cTbar^{\mu\a,kji}\ \G^\b_\a\ 
      Y_i^{\hskip 0.3em l}\ \cT_{\mu\b,ljk}
\nonumber\\
  \left( \cBbar\ \G\ \cB\ Y \right) &= 
    (-)^{(\eta_i+\eta_j)(\eta_k+\eta_l)}
      \cBbar^{\a,kji}\ \G^\b_\a\ 
      Y_k^{\hskip 0.30em l}\ \cB_{\b,ijl}\ ,\nonumber\\
  \left( \cBbar\ \G\ Y^\mu\ \cT_\mu \right) &= 
    \cBbar^{\a,kji}\ \G^\b_\a\ 
      \left(Y^\mu \right)_i^{\hskip 0.30em l}\
        \cT_{\mu\b,ljk}.
\end{align}
The leading order PQ Lagrangian is given by
\begin{eqnarray}
  \hL^{PQ}_{LO} &=&
    \left( \cBbar\ i\vD\ \cB \right)\ 
    +\ 2\am \left( \cBbar \cB \cM_+ \right)\ 
    +\ 2\bm \left( \cBbar \cM_+ \cB \right)\ 
    +\ 2\smb \left( \cBbar \cB \right) {\rm str}(\cM_+)
\nonumber\\
 && -
    \left(\cTbar^{\mu}\left[\ i\vD-\Delta\right]\cT_\mu \right)\ 
    +\ 2\gm\left(\cTbar^\mu \cM_+ \cT_\mu \right)\ 
    -\ 2\smt\left(\cTbar^\mu \cT_\mu \right) {\rm str}(\cM_+)
\nonumber\\
 && +
    2\a\left(\cBbar S^\mu \cB A_\mu \right)\ 
    +\ 2\beta \left(\cBbar S^\mu A_\mu \cB \right)\ 
    +\ 2{\cal H} \left( \cTbar^\nu S^\mu A_\mu \cT_\nu \right)
\nonumber\\
 && +
    \sqrt{\frac{3}{2}}{\C} \left[ \left( \cTbar^\nu A_\nu \cB
      \right)\ 
    +\ \left( \cBbar A_\nu \cT^\nu \right) \right]
\label{eq:leadlagPQ}\, .
\end{eqnarray} 
The fixed coefficient and higher dimensional operators Lagrangians are
given by
\begin{eqnarray}
  \hL^{PQ}_\frac{1}{M_B} &=& 
    -\left( \cBbar \frac{D_\perp^2}{2 M_B} \cB \right)\ 
      -\ \a \left( \cBbar \frac{S\cdot i \roarrow D}{M_B} \cB
      \vit\cdot A \right)\ 
    +\ \a \left( \cBbar \frac{i \loarrow D \cdot S}{M_B} \cB
      \vit\cdot A \right)
\nonumber\\
  && - \b \left( \cBbar \vit\cdot A \frac{S\cdot i \roarrow D}
       {M_B} \cB \right)\ 
     +\ \b \left( \cBbar \frac{i \loarrow D \cdot S}{M_B}
       \vit\cdot A \cB \right)
     +\ \left(\cTbar^\mu \frac{D_\perp^2}{2 M_B} \cT_\mu \right)
\label{eq:fixedPQ}\, ,
\end{eqnarray}
and~\footnote{It is interesting to note operators of the form
  $\big( \Bbar [A_\mu,A_\nu] [S^\mu,S^\nu] B \big)$ that vanish in
  $\chi$PT don't vanish in PQ$\chi$PT due to grading factors.
  However, they don't contribute to the octet baryon masses.}
\begin{eqnarray}
     \hL^{(2,4)}_{c.t.} &=& \frac{1}{( 4\pi f)}\Bigg\{
       \ \ b^A_1\, \Bbar^{kji} \left( A\cdot A \right)_i^{\hskip0.3em n} 
               B_{njk}\ 
        +\ b^A_2\, (-)^{(\eta_i+\eta_j)(\eta_k+\eta_n)} \Bbar^{kji}
           \left( A\cdot A \right)_k^{\hskip0.3em n} B_{ijn} \nonumber\\
       && \phantom{4\pi f an}
       \ + b^A_3\, (-)^{\eta_l(\eta_j+\eta_n)} \Bbar^{kji} (A_\mu
             )_i^{\hskip0.3em l} (A^\mu )_j^{\hskip0.3em n} B_{lnk}\ 
        +\ b^A_4\, \Bbar^{kji} B_{ijk} {\rm Tr}\left( A\cdot A
          \right) \nonumber\\
       && \phantom{4\pi f an}
       \ + b^{vA}_1\, \Bbar^{kji} \left( v\cdot\, A\, v\cdot A
           \right)_i^{\hskip0.3em n} B_{njk}\ 
        +\ b^{vA}_2\, (-)^{(\eta_i+\eta_j)(\eta_k+\eta_n)} \Bbar^{kji}
           \left( v\cdot A\, v\cdot A \right)_k^{\hskip0.3emn} B_{ijn} \nonumber\\ 
       && \phantom{4\pi f an}
       \ + b^{vA}_3\, (-)^{\eta_l(\eta_j+\eta_n)} \Bbar^{kji} (v\cdot
             A )_i^{\hskip0.3em l} (v\cdot A )_j^{\hskip0.3em n} B_{lnk}\ 
        +\ b^{vA}_4\, \Bbar^{kji} B_{ijk} {\rm Tr}\left( v\cdot A\, v\cdot A
          \right) \nonumber\\
       && \phantom{4\pi f an}
         \ + b^M_1\, \cBbar^{kji} \left( \cM_+ \cM_+ \right)_i^{\hskip0.3em n} 
               \cB_{njk}\ \ 
       +\ \ b^M_2\, (-)^{(\eta_i+\eta_j)(\eta_k+\eta_n)}
               \cBbar^{kji} \left( \cM_+ \cM_+ \right)_k^{\hskip0.3em n} 
               \cB_{ijn} \nonumber\\
       && \phantom{4\pi f an}
        \ + b^M_3\, (-)^{\eta_l(\eta_j+\eta_n)}
               \cBbar^{kji} (\cM_+)_i^{\hskip0.3em l}
               (\cM_+)_j^{\hskip0.3em n} \cB_{lnk}\ \ 
       +\ \ b^M_4\, \cBbar^{kji} \cB_{ijk}\ {\rm str}\left(\cM_+
               \cM_+\right) \nonumber\\
       && \phantom{4\pi f an}
        \ + b^M_5\, \cBbar^{kji} (\cM_+)_i^{\hskip0.3em n} \cB_{njk}\ 
               {\rm str}\left( \cM_+ \right)\ \ 
      + \ \ b^M_7\, \cBbar^{kji} \cB_{ijk}\ {\rm str}\left( \cM_+ \right)\
               {\rm str}\left( \cM_+ \right)  \nonumber\\
       && \phantom{4\pi f an}
        \ + b^M_6\, (-)^{(\eta_i+\eta_j)(\eta_k+\eta_n)}
               \cBbar^{kji} (\cM_+)_k^{\hskip0.3em n} \cB_{ijn}\ 
               {\rm str}\left( \cM_+ \right) \bigg\}
\label{eq:hdoPQ}\, .
\end{eqnarray}
All the coefficients appearing in Eq.~(\ref{eq:leadlagPQ}),
Eq.~(\ref{eq:fixedPQ}) and Eq.~(\ref{eq:hdoPQ}), have the same
numerical values as in $\chi$PT.  The situation is similar to the
$\chi$PT case considered in Section~\ref{s:HBXPTB}.  All the operators
in the Lagrangian can be multiplied by arbitrary functions of $\D /
\L_\chi$.  We include these effects by treating all the coefficients
in the Lagrangian as arbitrary polynomial functions of $\D / \L_\chi$,
and expand out to the appropriate order (see
Section~\ref{s:XPTmasses}).  All the coefficients appearing in this
expansion also have the same numerical values as in the $\chi$PT case.
%
%
%
%
%%%%%%%%%  PQXPT Masses  %%%%%%%%%%%%
%
%
%
%
\section{Baryon Masses in PQ$\chi$PT}
\begin{figure}[tb]
  \centering
  \includegraphics[width=0.5\textwidth]{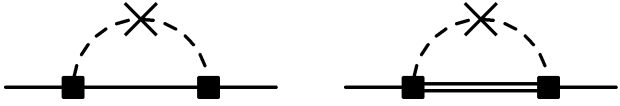}%
  \caption{
    In addition to the one-loop diagrams in Fig.~\ref{fig:NLO},
    $M_{B_i}^{(3/2)}$ also receives contributions from the singlet
    (hairpins) in PQ$\chi$PT.  Single and double lines correspond
    to $\mathbf {240}$-baryons and $\mathbf {138}$-baryons
    respectively.  The crossed 
    dashed line denotes a hairpin propagator.  The filled squares
    denote the axial coupling given in Eq.~(\ref{eq:leadlagPQ}).
  }
  \label{fig:PQNLO}
\end{figure}

\begin{figure}[tb]
  \centering
  \includegraphics[width=0.75\textwidth]{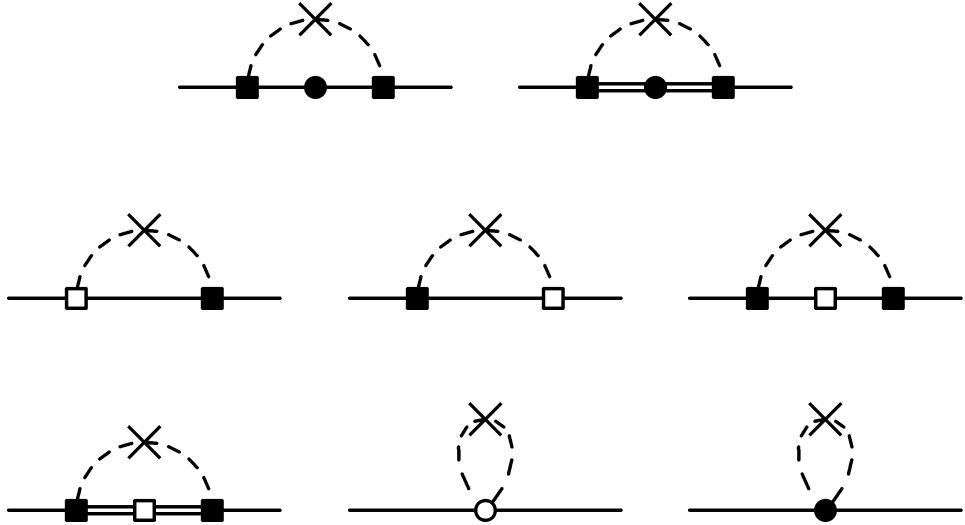}%
  \caption{
    In addition to the one-loop diagrams in Fig.~\ref{fig:NNLO},
    $M_{B_i}^2$ also receives contributions from the singlet
    (hairpins) in PQ$\chi$PT.  Single and double lines correspond
    to $\mathbf {240}$-baryons and $\mathbf {138}$-baryons
    respectively.  The crossed dashed line denotes a hairpin
    insertion.  Filled squares denote the axial coupling given in
    Eq.~(\ref{eq:leadlagPQ}).  Filled circles denote a coupling to
    $\cM_+$ given in Eq.~(\ref{eq:leadlagPQ}).  Empty squares and
    empty circles denote insertions of fixed $1/M_B$ operators given in
    Eq.~(\ref{eq:fixedPQ}), and insertions of a two-baryon-two-axial
    coupling given in Eq.~(\ref{eq:hdoPQ}) respectively.
  }
  \label{fig:PQNNLO}
\end{figure}

The chiral expansion of the octet baryon masses in PQ$\chi$PT has
the same form as in $\chi$PT.
\begin{eqnarray}
     M_{B_i} = M_0\left(\mu \right) \ -\ M_{B_i}^{(1)}\left(\mu \right)
               \ -\ M_{B_i}^{(3/2)}\left(\mu \right)
               \ -\ M_{B_i}^{(2)}\left(\mu \right)\ +\ .\, .\ .
\label{eq:massexpPQ}
\end{eqnarray}
However, we must also include hairpin contributions from the eta
field propagators.
See Fig.~\ref{fig:PQNLO}
and Fig.~\ref{fig:PQNNLO}.

We find that the mass expansion of the nucleon is:
\begin{eqnarray}\label{eq:NmassLOPQ}
  M_N^{(1)} &=& 2\mbar (\am+\bm)\ +\ 2\smb(2\mj+\mr)\\
\label{eq:NmassNLOPQ}
  M_N^{(3/2)} &=& \frac{1}{8 \pi f^2} \bigg[
    4 D(F-\frac{1}{3}D) \mp^3 +\frac{1}{3}(5D^2 -6DF + 9F^2)(2\mju^3 +
      \mru^3) \nonumber\\
  &&\qquad\quad
    +(D-3F)^2 \cM^3 (\mp,\mp) +\frac{2\C^2}{3\pi}\cF (\mp,\D,\mu)
    \nonumber\\
  &&\qquad\quad
    +\frac{2\C^2}{3\pi}\cF (\mju,\D,\mu) +\frac{C^2}{3}\cF
    (\mru,\D,\mu) \bigg],
\end{eqnarray}
consistent with \cite{Chen:2001yi}.  The contribution of this work is
the NNLO mass contribution to the octet baryon masses.
%
%%%%% Nucleon m_Q^2
%
\begin{eqnarray*}
  -M_N^{(2)} &=&
   \overline \cL (\mp,\mu) \bigg\{
     \frac{1}{(4\pi f)^3}\bigg[
      \frac{1}{2}b^A_3 +\frac{1}{8}b^{vA}_3 \bigg]
     -\frac{1}{M_B(4\pi f)^2} \bigg[
       \frac{3D(3F-D)}{2} +\frac{5\C^2}{4} \bigg] \bigg\} \nonumber\\
 && -\overline \cL (\mp,\mp,\mu) \bigg\{
      \frac{1}{(4\pi f)^3} \bigg[
        b^A_1 +b^A_2 +b^A_3 +\frac{1}{4}b^{vA}_1 +\frac{1}{4}b^{vA}_2
        +\frac{1}{4}b^{vA}_3 \bigg] 
        +\frac{9 (D-3F)^2}{8 M_B (4\pi f)^2} \bigg\}
        \nonumber\\
  &&-\overline \cL (\mju,\mu) \bigg\{
     \frac{1}{(4\pi f)^3} \bigg[
      2b^A_1 +2b^A_2 +\frac{1}{2}b^{vA}_1 +\frac{1}{2}b^{vA}_2 \bigg]
      \nonumber\\
  &&\qquad\qquad\qquad
     +\frac{1}{M_B(4\pi f)^2} \bigg[
       \frac{3(5D^2-6DF+9F^2)}{4} +\frac{5\C^2}{4} \bigg]
        \bigg\} \nonumber\\
  && -\overline \cL (\mru,\mu) \bigg\{ 
       \frac{1}{(4\pi f)^3} \bigg[
        b^A_1 +b^A_2 +\frac{1}{4}b^{vA}_1 +\frac{1}{4}b^{vA}_2 \bigg]
        \nonumber\\
  &&\qquad\qquad\qquad
       +\frac{1}{M_B (4\pi f)^2} \bigg[
        \frac{3}{8}(5D^2-6DF-9F^2) +\frac{5}{8}C^2 \bigg] \bigg\}
        \nonumber\\
  && -\frac{1}{(4\pi f)^3} \bigg[ b^A_4 +\frac{1}{4}b^{vA}_4 \bigg] \bigg\{
       4\overline \cL (\mjj,\mu) +4\overline \cL (\mjr,\mu) +\overline
       \cL (\mrr,\mu) \nonumber\\
  &&\qquad\qquad\qquad\qquad\qquad\qquad
     +2\overline \cL (\mjj,\mjj,\mu) +\overline \cL
       (\mrr,\mrr,\mu) \bigg\} \nonumber\\
  && +\cL (\mp,\mp,\mu) \frac{4\mbar (\am +\bm)}{(4\pi f)^2}
     +\frac{8\mj \smb}{(4\pi f)^2} \bigg[ 2\cL (\mjj,\mu) +\cL
       (\mjj,\mjj,\mu) \bigg] \nonumber\\
  && +\frac{4\smb}{(4\pi f)^2} \bigg\{
       2(\mj +\mr) \cL (\mjr,\mu)
       +\mr \bigg[ \cL (\mrr,\mu) +\cL (\mrr,\mrr,\mu) \bigg] \bigg\}
       \nonumber\\
  && +\cL (\mju,\mu) \frac{1}{(4\pi f)^2} \bigg\{
       4(\mbar+ \mj)(\am+\bm) +2(5D^2-6DF+9F^2) \MN \nonumber\\
  &&\qquad\qquad\qquad\qquad\qquad
       -9(D-F)^2 \MsSj -(D+3F)^2 \MsLj \bigg\} \nonumber\\
  && +\cL (\mru,\mu) \frac{1}{(4\pi f)^2} \bigg\{
       2(\mbar + \mr)(\am+\bm) +(5D^2-6DF+9F^2) \MN \nonumber\\
  &&\qquad\qquad\qquad\qquad\qquad
       -\frac{9}{2}(D-F)^2 \MsSr -\frac{1}{2}(D+3F)^2 \MsLr \bigg\}
       \nonumber\\
  && +\cJ (\mp,\mu) \frac{2C^2}{(4\pi f)^2} \bigg[
       \MN +\MD \bigg] 
     +\cJ (\mju,\mu) \frac{2C^2}{(4\pi f)^2} \bigg[
       \MN +\MsSSj \bigg] \nonumber\\
  && +\cJ (\mru,\mu) \frac{C^2}{(4\pi f)^2} \bigg[
       \MN +\MsSSr \bigg] 
       +\frac{b^{vA}_4}{2(4\pi f)^3} \bigg[
       \mjj^4 +\mjr^4 +\frac{1}{4}\mrr^4 \bigg] \nonumber\\ 
  && -\mp^4 \bigg\{
        \frac{1}{(4\pi f)^3} \frac{b^{vA}_3}{16}
        +\frac{1}{M_B(4\pi f)^2} \bigg[ \frac{5D(3F-D)}{4}
         +\frac{9C^2}{8} \bigg]\bigg\}\nonumber\\
  && +\mju^4 \bigg\{
        \frac{b^{vA}_1 +b^{vA}_2}{4(4\pi f)^3} 
        -\frac{5(5D^2-6DF+9F^2) +9C^2}{8M_B(4\pi f)^2} \bigg\} \nonumber\\
  && +\mru^4 \bigg\{
        \frac{b^{vA}_1 +b^{vA}_2}{8(4\pi f)^3} 
        -\frac{5(5D^2-6DF+9F^2) +9C^2}{16M_B(4\pi f)^2} \bigg\}
        +\rm {cont.} \nonumber\\
\end{eqnarray*}
\begin{eqnarray}\label{eq:NmassNNLOPQ}
  && +\cM^4 (\mp,\mp) \bigg\{
        \frac{b^{vA}_1 +b^{vA}_2 +b^{vA}_3}{8(4\pi f)^3} 
        -\frac{15(D-3F)^2}{16M_B(4\pi f)^2} \bigg\}\nonumber\\
  && +\frac{b^{vA}_4}{8(4\pi f)^3} \bigg[ 2\cM^4 (\mjj,\mjj) + \cM^4
        (\mrr,\mrr) \bigg] \nonumber\\
  && +\mp^2 \frac{2}{(4\pi f)^2} \bigg[ \MN +\MD \bigg] \nonumber\\
  && +\mju^2 \frac{2}{(4\pi f)^2} \bigg\{
       \frac{2}{3}(5D^2-6DF+9F^2)\MN -3(D-F)^2\MsSj \nonumber\\
  &&\qquad\qquad\qquad\quad
      -\frac{1}{3}(D+3F)^2\MsLj +C^2 \Big[ \MN +\MsSSj \Big] \bigg\}\nonumber\\
  && +\mru^2 \frac{1}{(4\pi f)^2} \bigg\{
       \frac{2}{3}(5D^2-6DF+9F^2)\MN -3(D-F)^2\MsSr \nonumber\\
  &&\qquad\qquad\qquad\quad
      -\frac{1}{3}(D+3F)^2\MsLr +C^2 \Big[ \MN +\MsSSr \Big] \bigg\}\nonumber\\
  && +\frac{\mbar^2}{(4\pi f)} (b^M_1 +b^M_2 +b^M_3)
     +\frac{\mbar \mj}{(4\pi f)} (2b^M_5 +2b^M_6) 
     +\frac{\mj^2}{(4\pi f)} (2b^M_4 +4b^M_7) \nonumber\\
  && +\frac{\mbar \mr}{(4\pi f)} (b^M_5 +b^M_6) 
     +\frac{\mj \mr}{(4\pi f)} 4b^M_7
     +\frac{\mr^2}{(4\pi f)} (b^M_4 +b^M_7),
\end{eqnarray}
where the functions not already defined in Eq.~(\ref{eq:Xlogs1}) are,
\begin{eqnarray}
  \cM^n (\mphi,m_{\phi^\prime}) &=& {\cal H}_{\phi \phi^\prime} 
    (\mphi^n,m_{\phi^\prime}^n,\mX^n)\nonumber\\
  \cL (\mphi,m_{\phi^\prime}) &=& {\cal H}_{\phi \phi^\prime}
    \Big( \cL (\mphi,\mu) ,\cL (m_{\phi^\prime},\mu) ,\cL (m_X,\mu)
     \Big) \nonumber\\
  \overline \cL (\mphi,m_{\phi^\prime}) &=& {\cal H}_{\phi \phi^\prime}  
     \Big( \overline \cL (\mphi,\mu) ,
           \overline \cL (m_{\phi^\prime},\mu) ,
           \overline \cL (m_X,\mu) \Big).
\end{eqnarray}
We stress that these mass expressions,
Eq.~(\ref{eq:NmassLOPQ})-(\ref{eq:NmassNNLOPQ}), are quark mass
expansions.  If a meson mass expansion is desired, one can express
these quark masses in terms of the lattice meson masses.
To do this consistently, one must replace the quark masses in
Eq.~(\ref{eq:NmassLOPQ}) with the NLO relation to the PQ$\chi$PT meson
masses, while the quark masses in Eq.~(\ref{eq:NmassNLOPQ}) and
Eq.~(\ref{eq:NmassNNLOPQ}) can be replaced by the LO relation to the
PQ meson masses.  It is easy to show that these expressions reduce to
the $\chi$PT expressions, Eq.~(\ref{eq:NmassLO})-(\ref{eq:NmassNNLO}),
in the limit that $\mj \rightarrow \mbar$ and $\mr \rightarrow \ms$.
The results for the NNLO mass calculations for the remaining octet
baryons can be found in Appendix~\ref{ap:PQmasses}.

\section{Conclusions}
  We have calculated the $\Or (m_q^2)$ contribution to the masses of the
octet baryons in the isospin limit of $\chi$PT and PQ$\chi$PT,
keeping the decuplet baryons as dynamical intermediate states.
Working to this order in the chiral expansion introduces a large
number of LEC's.  In order for the calculation to have any predictive
power, these LEC's must be fit from various experiments and or lattice
results.  Lattice calculations will eventually allow first principles
determination of these constants, and thus predictions of QCD
observables.  However, for the time being, to make rigorous
predictions of the baryon mass spectra arising from QCD, one needs to
perform PQQCD lattice simulations and extrapolate these results using
PQ$\chi$PT.

To date, there are limited PQQCD lattice calculations of the octet baryon
masses.  We hope this work will help to
further interest an execution of 
these calculations.  This is the first calculation
to $\Or (m_q^2)$ in the baryon sector using PQ$\chi$PT.  Calculations
of this order are necessary for reducing the uncertainty in the chiral
extrapolations and systematically studying the chiral expansion of
both PQ$\chi$PT and $\chi$PT.

\bigskip

\emph{Note Added.}
While this work was being completed, an independent analysis by Frink
and Meissner appeared~\cite{Frink:2004ic}.  Their work considers the
octet baryon masses in $\chi$PT only.

\acknowledgements
We would especially like to thank Martin Savage and Brian Tiburzi for
many useful discussions and encouragement.  
We would also like to thank Daniel Arndt, Will Detmold,
David Lin, and Ruth Van de Water for their input and
help with this manuscript.  
%And we would like to thank Erin Warnock, at the very least for all
%the coffee and cookies, and of course the motivational sass.  
This work was supported in part by the U.S. Department of
Energy under Grant No. DE-FG03-97ER4014.

%%%%%%%%%%%% Appendix
%
%
%
%
%
%
%
\appendix
\section{Chiral Logs and other Functions}\label{ap:Xlogs}
For the readers convenience, here we list all the functions arising in
the calculation of the octet baryon masses, including the previously
defined ones (which have been left unnumbered).
\begin{eqnarray*}
  \cL (\mphi,\mu) &=& 
    \mphi^2\ \ln \left( \frac{\mphi^2}{\mu^2} \right) \notag\\
  \overline \cL (\mphi,\mu) &=& 
    \mphi^4 \ \ln \left( \frac{\mphi^2}{\mu^2} \right) \notag\\
  \cF (\mphi,\D,\mu) &=& 
    \left( \mphi^2 - \D^2 \right) \left[ \sqrt{\D^2 - \mphi^2}\ 
      \ln \left( \frac{\D - \sqrt{\D^2 - \mphi^2 + i\e}} 
      {\D + \sqrt{\D^2 - \mphi^2 + i\e}} \right) - 
      \D \ln \left(\frac{\mphi^2}{\mu^2}\right) \right]\nonumber\\
  && \qquad\qquad\qquad\qquad
    -\frac{1}{2}\D \mphi^2 \ln \left(\frac{\mphi^2}{\mu^2}
      \right) \notag\\
  \cJ \left(\mphi,\D,\mu \right) &=&
    \left( \mphi^2 - 2\D^2 \right) \ln \left(\frac{\mphi^2}
      {\mu^2}\right) \nonumber\\
  && \qquad\qquad
    +2\D \sqrt{\D^2 -\mphi^2}\ \ln \left( \frac{\D - \sqrt{\D^2 -
          \mphi^2 +i\e}} {\D + \sqrt{\D^2 - \mphi^2 + i\e}} \right) \notag\\
\end{eqnarray*}
\begin{eqnarray}
  \cM^n (\mphi,m_{\phi^\prime}) &=& 
    {\cal H}_{\phi \phi^\prime} (\mphi^n,m_{\phi^\prime}^n,\mX^n) \nonumber\\
  \cL (\mphi,m_{\phi^\prime}) &=& 
    {\cal H}_{\phi \phi^\prime} \Big( \cL (\mphi,\mu) ,\cL
      (m_{\phi^\prime},\mu) ,\cL (m_X,\mu) \Big) \nonumber\\
  \overline \cL (\mphi,m_{\phi^\prime}) &=& 
    {\cal H}_{\phi \phi^\prime} \Big( \overline \cL (\mphi,\mu),
      \overline \cL (m_{\phi^\prime},\mu),
      \overline \cL (m_X,\mu) \Big) \notag\\
  \cF (\mphi,m_{\phi^\prime}) &=& 
    {\cal H}_{\phi \phi^\prime} \Big( \cF (\mphi,\D,\mu), \cF
      (m_{\phi^\prime},\D,\mu), \cF (\mX,\D,\mu) \Big)\\
  \cJ (\mphi,m_{\phi^\prime}) &=&
    {\cal H}_{\phi \phi^\prime} \Big( \cJ (\mphi,\D,\mu),
      \cJ (m_{\phi^\prime},\D,\mu), \cJ (\mX,\D,\mu) \Big)
\end{eqnarray}
And ${\cal H}_{ab}(A,B,C)$ is defined in Eq.~(\ref{eq:Hfunction}).
\section{Octet Baryon Masses in $\chi$PT}\label{ap:XPTmasses}

The mass of the $i^{th}$ baryon in the chiral expansion is
\begin{eqnarray}
  M_{B_i} = M_0\left(\mu \right) \ -\ M_{B_i}^{(1)}\left(\mu \right)
            \ -\ M_{B_i}^{(3/2)}\left(\mu \right)
            \ -\ M_{B_i}^{(2)}\left(\mu \right)\ +\ .\, .\ .
\end{eqnarray}
The LO and NLO mass corrections to the octet baryons are listed for
completeness.  The LO mass corrections are listed in
Table~\ref{t:MLO}.  The NLO mass corrections are given by,
\begin{equation}
  -\ M_{B_i}^{(3/2)} =
     \frac{-2}{(4\pi f)^2} \sum_\phi \biggl[
     (C_{BB\phi})^2 \pi \mphi^3\nonumber
     + C^2 (C_{TB\phi})^2 \cF (\mphi,\D,\mu) \biggr],
\end{equation}
The sum on $\phi$ is over loop mesons with mass $\mphi$.  The
$C_{BB\phi}$ and $C_{TB\phi}$ are Clebsch-Gordon coefficients between
the Axial field and two octet baryons or an octet and decuplet
baryon.  The sum of these coefficients are listed in
Table~\ref{t:BBATBA}.

Results of this work are the NNLO mass corrections in $\chi$PT to the octet
baryons given by the expression,
\begin{eqnarray}
  -\ M_{B_i}^{(2)} &=&
% b^A's
     -\frac{1}{(4 \pi f)^3} \sum_\phi \left[
        C^A_{BB\phi\phi} +\frac{1}{4} C^{vA}_{BB\phi\phi}
         \right] \overline \cL (\mphi,\mu) 
     + \frac{1}{(4 \pi f)^3} \sum_\phi %\left[
         C^{vA}_{BB\phi\phi} \frac{1}{8} 
          \mphi^4 \nonumber\\
% Kinetic Corrections
     && - \frac{1}{(4\pi f)^2} \frac{3}{8 M_B} \sum_\phi \bigg\{
            (C_{BB\phi})^2 \left[ 3 \overline \cL (\mphi,\mu)
              +\frac{5}{2}\mphi^4 \right] \nonumber\\
% Kinetic Decuplet Correction
     &&\qquad\qquad\qquad\qquad\qquad
        + C^2 (C_{TB\phi})^2 \left[ 5\overline \cL (\mphi,\mu)
          +\frac{9}{2}\mphi^4 \right] \bigg\} \nonumber\\
% Internal Octet Mass insertion and Wavefunction Correction
     && + \frac{3}{(4\pi f)^2} \bigg\{ 
          M^{(1)}_{B_i} \sum_\phi (C_{BB\phi})^2\ \bigg[ \cL (\mphi,\mu)
          +\frac{2}{3} \mphi^2 \bigg] \nonumber\\
     &&\qquad\qquad\qquad\qquad\qquad
          - \sum_{\phi,j} (C_{BB\phi})^2\ M_{B_j}^{(1)}\ \left[ \cL
            (\mphi,\mu) +\frac{2}{3} \mphi^2 \right] \bigg\} \nonumber\\
% Internal Decuplet Mass Insertion and Wave Function Correction
     && + \frac{3 C^2}{(4\pi f)^2} \bigg\{
            M^{(1)}_{B_i} \sum_\phi (C_{TB\phi})^2 \bigg[ \cJ
             \left(\mphi,\D,\mu \right) +\mphi^2 \bigg] \nonumber\\
     &&\qquad\qquad\qquad\qquad\qquad
          + \sum_{\phi,j} (C_{TB\phi})^2 M^{(1)}_{T_j}  \bigg[ \cJ
          \left(\mphi,\D,\mu \right) +\mphi^2 \bigg] \bigg\} \nonumber\\
% M_+ to 2nd order
     && + \frac{1}{(4\pi f)^2} \sum_\phi C^M_{BB\phi\phi}\
            \cL (\mphi,\mu)
% M_+ M_+ insertion
      - \frac{1}{(4 \pi f)} \sum_{i,j} C_{B_{ij}} m_i m_j
\end{eqnarray}
The coefficients for the masses are listed in Table~\ref{t:BBATBA}
through Table~\ref{t:MM}.

%
%
%      XPT Tables
%
%
%
\begin{table}[!h]
\caption{\label{t:MLO}
The LO Octet and Decuplet Baryon Masses in $\chi$PT.}
\begin{ruledtabular}
\begin{tabular}{c | c | c | c }
  & Octet Baryons   & &   Decuplet Baryons    \\
  & $M_B^{(1)}$ & &   $M_T^{(1)}$        \\
  \hline
    $\MN$ & $2\mbar (\am +\bm +2\smb) +2\ms \smb$
  & $\MD$ & $2\mbar(\gm -2\smt) -2\ms\smt$ \\
  \hline
    $\MS$ & $\mbar \left(\frac{5}{3}\am +\frac{2}{3}\bm  +4\smb \right)$
  & $\MSS$ & $\frac{2}{3}(2\mbar +\ms)(\gm -3\smt)$ \\
    & $+\ms \left(\frac{1}{3}\am +\frac{4}{3}\bm +2\smb \right)$ & \\
  \hline
    $\ML$ & $\mbar (\am +2\bm +4\smb) +\ms(\am +2\smb)$
  & $\MXS$& $\frac{2}{3}\mbar(\gm -6\smt) +\frac{2}{3}\ms(2\gm
     -3\smt)$ \\
  \hline
    $\MX$ & $\mbar \left(\frac{1}{3}\am +\frac{4}{3}\bm +4\smb \right)$
  & $\MO$ & $2\ms(\gm -\smt) -4\smt \mbar$ \\
    & $ +\ms \left(\frac{5}{3}\am +\frac{2}{3}\bm +2\smb \right)$ & \\
\end{tabular}
\end{ruledtabular}
\end{table} 

\begin{table}[!h]
\caption{\label{t:BBATBA}
  The octet-octet-axial and decuplet-octet-axial coupling coefficients
in $\chi$PT.  The coefficients have been grouped according to loop
mesons with mass $\mphi$ for each octet baryon.}
\begin{ruledtabular}
\begin{tabular}{c | c c c | c c c}
   & & $\sum_\phi (C_{BB\phi})^2$ & & & $\sum_\phi (C_{TB\phi})^2$ \\
       $\phi$ 
     & $\pi$ 
     & $\eta$ 
     & $K$ 
     & $\pi$ 
     & $\eta$ 
     & $K$  \\
   \hline
% Nucleon
   $N$ & $\frac{3}{2}(D+F)^2 $ 
       & $\frac{1}{6}(D-3F)^2 $ 
       & $\frac{1}{3}(5D^2 -6DF +9F^2)$ 
       & $\frac{4}{3}$ 
       & $0$ 
       & $\frac{1}{3}$ \\ \hline
% Sigma
   $\S$ & $\frac{2}{3} (D^2 +6F^2)$ 
        & $\frac{2}{3}D^2$ 
        & $2(D^2 +F^2)$ 
        & $\frac{2}{9}$ 
        & $\frac{1}{3}$ 
        & $\frac{10}{9}$ \\ \hline
% Lambda
   $\L$ & $2 D^2$ 
        & $\frac{2}{3} D^2$ 
        & $\frac{2}{3}(D^2 + 9F^2)$
        & $1$ 
        & $0$ 
        & $\frac{2}{3}$ \\ \hline
% Cascade
   $\X$ & $\frac{3}{2} (D-F)^2$ 
        & $\frac{1}{6} (D+3F)^2$ 
        & $\frac{1}{3}(5D^2 +6DF +F^2)$ 
        & $\frac{1}{3}$ 
        & $\frac{1}{3}$
        & $1$ \\
\end{tabular}
\end{ruledtabular}
\end{table}
%%%%%%%%%%%%%%%  Ahdo Coefficients %%%%%%
\begin{table}[!h]
\caption{\label{t:BBAA}
  The coefficients of 2-octet 2-axial couplings in $\chi$PT.  The
  coefficients have been grouped according to loop mesons with mass
  $\mphi$ for each octet baryon.}
\begin{ruledtabular}
\begin{tabular}{c| c c c}
   & & $\sum_\phi C^A_{BB\phi\phi}$ and $\sum_\phi C^{vA}_{BB\phi\phi}$ \\
     & $\pi$ 
     & $\eta$ 
     & $K$ \\
   \hline
% Nucleon
   $N$ & $\frac{3}{2}b^A_1 +\frac{3}{2}b^A_2 -b^A_3 +3b^A_4$ 
       & $\frac{1}{6}b^A_1 +\frac{1}{6}b^A_2 +\frac{1}{6}b^A_3 +b^A_4$
       & $b^A_1 +b^A_2 +4b^A_4$\\ \hline
% Sigma
   $\S$ & $\frac{1}{2}b^A_1 +\frac{5}{4}b^A_2 +\frac{1}{12}b^A_3
          +3b^A_4$ 
        & $\frac{1}{2}b^A_1 +\frac{1}{4}b^A_2 -\frac{1}{4}b^A_3
          +b^A_4$ 
        & $\frac{5}{3}b^A_1 +\frac{7}{6}b^A_2 -\frac{2}{3}b^A_3
          +4b^A_4$ \\ \hline
% Lambda
   $\L$ & $\frac{3}{2}b^A_1 +\frac{3}{4}b^A_2 -\frac{3}{4}b^A_3
          +3b^A_4$ 
        & $\frac{1}{6}b^A_1 +\frac{5}{12}b^A_2 -\frac{1}{12}b^A_3
          +b^A_4$ 
        & $b^A_1 +\frac{3}{2}b^A_2 +4b^A_4$ \\ \hline
% Cascade
   $\X$ & $b^A_1 +\frac{1}{4}b^A_2 +3b^A_4$ 
        & $\frac{1}{3}b^A_1 +\frac{7}{12}b^A_2 -\frac{1}{6}b^A_3
          +b^A_4$ 
        & $\frac{4}{3}b^A_1 +\frac{11}{6}b^A_2 -\frac{2}{3}b^A_3
          +4b^A_4$ \\
\end{tabular}
\end{ruledtabular}
\end{table}
%%%%%%%%%%%%%%%  M+ 2nd order coefficients
\begin{table}[!h]
\caption{\label{t:MBBphiphi}
  The $C^M_{BB\phi\phi}$ coefficients in $\chi$PT.  The coefficients
  have been grouped according to loop mesons with mass $\mphi$ for
  each octet baryon.}
\begin{ruledtabular}
\begin{tabular}{c| c c c}
   & & $C^M_{BB\phi\phi}$ & \\
   & $\pi$ 
   & $\eta$ 
   & $K$ \\
   \hline
% Nucleon
   $N$ & $6 \mbar (\am +\bm +2 \smb)$ 
       & $\frac{2}{3} \mbar (\am +\bm +2\smb) +\frac{8}{3} m_s \smb$ 
       & $2 (\mbar +m_s)(\am +\bm +4\smb)$ \\ \hline
% Sigma
   $\S$ & $\mbar (5\am +2\bm +12\smb)$ 
        & $\mbar \left( \frac{5}{9}\am +\frac{2}{9}\bm
          +\frac{4}{3}\smb \right)$ 
        & $(\mbar +m_s) \left( \frac{7}{3}\am +\frac{10}{3}\bm +8\smb
          \right)$ \\
        & & $+m_s \left( \frac{4}{9}\am + \frac{16}{9}\bm
          +\frac{8}{3}\smb \right)$ \\
   \hline
% Lambda
   $\L$ & $3\mbar (\am +2\bm +4\smb)$ 
        & $\mbar \left( \frac{1}{3}\am +\frac{2}{3}\bm
          +\frac{4}{3}\smb \right)$ 
        & $(\mbar +m_s)(3\am +2\bm +8\smb)$  \\
        & & $+m_s \left( \frac{4}{3}\am +\frac{8}{3}\smb \right)$ \\ \hline
% Cascade
   $\X$ & $\mbar (\am +4\bm +12\smb)$ 
        & $\mbar \left( \frac{1}{9}\am +\frac{4}{9}\bm
          +\frac{4}{3}\smb \right)$ 
        & $(\mbar +m_s) \left( \frac{11}{3}\am +\frac{8}{3}\bm +8\smb
          \right)$ \\
        & & $+m_s \left( \frac{20}{9}\am +\frac{8}{9}\bm
          +\frac{8}{3}\smb \right)$ \\
\end{tabular}
\end{ruledtabular}
\end{table}
%%%%%%%%%%%%%%%  Internal mass insertion coefficients and Wavefunction
\begin{table}[!h]
\caption{\label{t:BmassInsert}
  The coefficients for the internal octet mass insertion and the
  wavefunction correction in $\chi$PT.  The coefficients
  have been grouped according to loop mesons with mass $\mphi$ for
  each octet baryon.}
\begin{ruledtabular}
\begin{tabular}{c| c c c}
   & & $M^{(1)}_{B_i} \sum_\phi (C_{BB\phi})^2
        - \sum_{\phi,j} (C_{BB\phi})^2 M^{(1)}_{B_j}$ & \\
   & $\pi$ 
   & $\eta$ 
   & $K$\\
   \hline
% Nucleon       
   $N$ & $0$ 
       & $0$ 
       & $\frac{1}{3}(5D^2-6DF+9F^2) \MN$
       \\
       & & & $-\frac{3}{2}(D-F)^2 \MS -\frac{1}{6}(D+3F)^2 \ML$ \\ \hline
% Sigma
   $\S$ & $\frac{2}{3}D^2 (\MS -\ML)$ 
        & $0$ 
        & $2(D^2+F^2)\MS$
        \\
        & & & $-(D-F)^2 \MN - (D+F)^2 \MX$\\\hline
% Lambda
   $\L$ & $2D^2 (\ML -\MS)$ 
        & $0$ 
        & $\frac{2}{3}(D^2+9F^2)\ML$
        \\
        & & & $-\frac{1}{3}(D+3F)^2 \MN -\frac{1}{3}(D-3F)^2 \MX$\\ \hline
% Cascade
   $\X$ & $0$ 
        & $0$ 
        & $\frac{1}{3}(5D^2+6DF+9F^2)\MX$
        \\
        & & & $-\frac{3}{2}(D+F)^2 \MS -\frac{1}{6} (D-3F)^2 \ML$ \\
\end{tabular}
\end{ruledtabular}
\end{table}
%%%%%%%%%%%%% Internal Decuplet Mass Insertion
\begin{table}[!h]
\caption{\label{t:TmassInsert}
  The coefficients for an internal decuplet mass insertion and the
  wavefunction correction in $\chi$PT.  The coefficients
  have been grouped according to loop mesons with mass $\mphi$ for
  each octet baryon.}
\begin{ruledtabular}
\begin{tabular}{c| c c c}
   & & $\sum_{\phi,j} (C_{TB\phi})^2 M^{(1)}_{T_j}
        +M^{(1)}_{B_i} \sum_\phi (C_{TB\phi})^2$ & \\
   & $\pi$ 
   & $\eta$ 
   & $K$ \\
   \hline
% Nucleon       
   $N$ & $\frac{4}{3} (\MD +\MN)$ 
       & $0$ 
       & $\frac{1}{3} (\MSS +\MN)$ \\ \hline
% Sigma
   $\S$ & $\frac{2}{9} (\MS +\MSS)$ 
        & $\frac{1}{3} (\MS +\MSS)$ 
        & $\frac{2}{9} (5\MS +4\MD +\MXS)$ \\ \hline
% Lambda
   $\L$ & $\ML +\MSS$ 
        & $0$ 
        & $\frac{2}{3} (\ML +\MXS)$ \\ \hline
% Cascade
   $\X$ & $\frac{1}{3} (\MX +\MXS)$ 
        & $\frac{1}{3} (\MX +\MXS)$ 
        & $\frac{1}{3} (3\MX +2\MO +\MSS)$ \\
\end{tabular}
\end{ruledtabular}
\end{table}
%%%%%%%%%%%%%%% M_+ M_+ Insertion
\begin{table}[!h]
\caption{\label{t:MM}
  The $C_{B_{ij}}$ coefficients in $\chi$PT.  The coefficients are
  grouped by products of quark masses for each octet baryon.}
\begin{ruledtabular}
\begin{tabular}{c| c c c}
   & & $\sum_{i,j} C_{B_{ij}} m_i m_j$ & \\
   & $\mbar^2$ 
   & $\mbar m_s$ 
   & $m_s^2$ \\
   \hline
% Nucleon
   $N$ & $b^M_1 +b^M_2 +b^M_3 +2b^M_4$ 
       & $b^M_5 +b^M_6 +4b^M_7$ 
       & $b^M_4 +b^M_7$ 
       \\
       & $+2b^M_5 +2b^M_6 +4b^M_7$ \\
   \hline
% Sigma
   $\S$ & $\frac{1}{3}b^M_1 +\frac{5}{6}b^M_2 +\frac{1}{6}b^M_3
           +2b^M_4$ 
        & $\frac{5}{6}b^M_3 + \frac{5}{3}b^M_5 +\frac{7}{6}b^M_6 +4b^M_7$
        & $\frac{2}{3}b^M_1 +\frac{1}{6}b^M_2 +b^M_4
          +\frac{2}{3}b^M_5$
        \\
        & $+\frac{2}{3}b^M_5 +\frac{5}{3}b^M_6 +4b^M_7$ 
        & 
        & $+\frac{1}{6}b^M_6 +b^M_7$ \\
   \hline
% Lambda
   $\L$ & $b^M_1 +\frac{1}{2}b^M_2 +\frac{1}{2}b^M_3 +2b^M_4$ 
        & $\frac{1}{2}b^M_3 +b^M_5 +\frac{3}{2}b^M_6 +4b^M_7$ 
        & $\frac{1}{2}b^M_2 +b^M_4 +\frac{1}{2}b^M_6 +b^M_7$ 
        \\
        & $+2b^M_5 +b^M_6 +4b^M_7$ \\
   \hline
% Cascade
   $\X$ & $\frac{2}{3}b^M_1 +\frac{1}{6}b^M_2 +2b^M_4
          +\frac{4}{3}b^M_5$ 
        & $\frac{5}{6}b^M_3 + \frac{4}{3}b^M_5 +\frac{11}{6}b^M_6
          +4b^M_7$ 
        & $\frac{1}{3}b^M_1 +\frac{5}{6}b^M_2 +\frac{1}{6}b^M_3
          +b^M_4$
        \\
        & $+\frac{1}{3}b^M_6 +4b^M_7$ 
        & & $+\frac{1}{3}b^M_5 +\frac{5}{6}b^M_6 +b^M_7$ \\
\end{tabular}
\end{ruledtabular}
\end{table}

\pagebreak
%
%
%
%   PQXPT masses
%
%
%
%
\section{Octet Baryon Masses in PQ$\chi$PT}
\label{ap:PQmasses}
The mass of the $i^{th}$ baryon in the chiral expansion of PQHB$\chi$PT
has the same form as in HB$\chi$PT.
\begin{eqnarray}
  M_{B_i} = M_0\left(\mu \right) \ -\ M_{B_i}^{(1)}\left(\mu \right)
            \ -\ M_{B_i}^{(3/2)}\left(\mu \right)
            \ -\ M_{B_i}^{(2)}\left(\mu \right)\ +\ .\, .\ .
\end{eqnarray}
The LO and NLO mass corrections to the octet baryons in PQ$\chi$PT are
listed here for completeness, and are consistent with
\cite{Chen:2001yi}.  The LO mass corrections for the $\mathbf {240}$-
and $\mathbf {138}$-baryons are listed in Table~\ref{t:VVVmasses},
Table~\ref{t:SGBmasses} and Table~\ref{t:SGTmasses}.  The NLO mass
corrections are,
\begin{eqnarray}
  -\ M_{B_i}^{(3/2)} &=&
     \frac{-2}{(4\pi f)^2} \sum_\phi \biggl[
       (C^{PQ}_{BB\phi})^2 \pi \mphi^3\nonumber
       + C^2 (C^{PQ}_{TB\phi})^2 \cF (\mphi,\D,\mu) \biggr] \nonumber\\
     && -\frac{2}{(4\pi f)^2} \sum_{\phi,\phi^\prime} \biggl[
       \pi\, {\overline C}_{B\phi \phi^\prime}\, \cM^3
       (\mphi,m_{\phi^\prime}) + C^2\, {\overline C}_{T\phi
         \phi^\prime} \cF (\mphi,m_{\phi^\prime}) \biggr].
\end{eqnarray}
The sum on $\phi$ is over loop mesons with mass $\mphi$.  The
$C^{PQ}_{BB\phi}$ and $C^{PQ}_{TB\phi}$ are PQHB$\chi$PT
Clebsch-Gordon coefficients between the axial field and the baryon
fields.  The sums on $\phi,\phi^\prime$ are over contributions from
the hairpin eta field interactions.  The sum is over pairs of eta
fields, with no double counting.

The main results of this work are the NNLO PQHB$\chi$PT octet mass
calculations.
\begin{eqnarray}
  -\ M_{B_i}^{(2)} &=&
% b^A's
     -\frac{1}{(4 \pi f)^3} \bigg\{
       \sum_\phi \left[ 
        C^{A,PQ}_{BB\phi\phi} +\frac{1}{4} C^{vA,PQ}_{BB\phi\phi}
        \right] \overline \cL (\mphi,\mu) \nonumber\\
     &&\qquad\qquad\qquad\qquad\qquad
      +\sum_{\phi \phi^\prime} \left[
        {\overline C}^A_{BB\phi\phi^\prime} +\frac{1}{4} {\overline
          C}^{vA}_{BB\phi\phi^\prime} \right] \overline \cL
      (\mphi,m_{\phi^\prime}) \bigg\}\nonumber\\
     && + \frac{1}{(4 \pi f)^3} \bigg\{ 
       \sum_\phi \frac{1}{8} C^{vA,PQ}_{BB\phi\phi} \mphi^4 
      +\sum_{\phi \phi^\prime} \frac{1}{8} {\overline
            C}^{vA}_{BB\phi\phi^\prime} \cM^4 (\mphi,m_{\phi^\prime})
          \bigg\} \nonumber\\
% Kinetic Corrections
     && - \frac{1}{M_B (4\pi f)^2} \frac{3}{8} \bigg\{ \sum_\phi 
            (C^{PQ}_{BB\phi})^2 \left[ 3\, \overline \cL (\mphi,\mu)
              +\frac{5}{2} \mphi^4 \right] \nonumber\\
     &&\qquad\qquad\qquad\qquad\qquad
        + \sum_{\phi \phi^\prime} {\overline C}_{B\phi \phi^\prime}
        \left[ 3 \overline \cL (\mphi,m_{\phi^\prime}) +\frac{5}{2}
          \cM^4 (\mphi,m_{\phi^\prime}) \right] \bigg\} \nonumber\\
% Kinetic Decuplet Correction
     && - \frac{C^2}{M_B (4\pi f)^2} \frac{3}{8} \bigg\{ \sum_\phi 
        (C^{PQ}_{TB\phi})^2 \left[ 5\, \overline \cL (\mphi,\mu)
          +\frac{9}{2}\mphi^4 \right] \nonumber\\
     &&\qquad\qquad\qquad\qquad\qquad
        + \sum_{\phi \phi^\prime} {\overline C}_{T\phi
          \phi^\prime} \left[ 5 \overline \cL (\mphi,m_{\phi^\prime})
          +\frac{9}{2} \cM^4 (\mphi,m_{\phi^\prime}) \right] \bigg\} \nonumber\\
% Wave function -Internal Octet Mass insertion 
     && + \frac{3}{(4\pi f)^2} \bigg\{ 
          M^{(1)}_{B_i} \sum_\phi (C^{PQ}_{BB\phi})^2\ \bigg[ \cL (\mphi,\mu)
          +\frac{2}{3} \mphi^2 \bigg] \nonumber\\
     &&\qquad\qquad\qquad\qquad\qquad
           -\sum_{j,\phi} M_{B_j}^{(1)} (C^{PQ}_{BB\phi})^2 \left[ \cL
             (\mphi,\mu) + \frac{2}{3} \mphi^2 \right] \bigg\} \nonumber\\
% Internal Decuplet mass insertion
     && + \frac{3C^2}{(4\pi f)^2} \bigg\{ 
           M^{(1)}_{B_i} \sum_\phi (C^{PQ}_{TB\phi})^2 \bigg[
          +\cJ \left(\mphi,\D,\mu \right) +\mphi^2 \bigg] \nonumber\\
     &&\qquad\qquad\qquad\qquad\qquad
           +  \sum_{j,\phi} M^{(1)}_{T_j} (C^{PQ}_{TB\phi})^2 \bigg[
          \cJ \left(\mphi,\D,\mu \right) + \mphi^2 \bigg]\bigg\} \nonumber\\
     && + \frac{3C^2}{(4\pi f)^2} \bigg\{ 
     M^{(1)}_{B_i} \sum_{\phi \phi^\prime} {\overline C}_{T\phi
       \phi^\prime} \bigg[ \cM^2 (\mphi,m_{\phi^\prime}) + \cJ
     (\mphi,m_{\phi^\prime}) \bigg] \nonumber\\
     &&\qquad\qquad\qquad\qquad\qquad
     +  \sum_{j,\phi\phi^\prime} M_{T_j}^{(1)}\ {\overline
          C}_{T\phi \phi^\prime} \bigg[ \cM^2 (\mphi,m_{\phi^\prime})
          + \cJ (\mphi,m_{\phi^\prime}) \bigg]\bigg\} \nonumber\\
% M_+ to 2nd order
     && + \frac{1}{(4\pi f)^2} \bigg\{ 
          \sum_\phi C^{M,PQ}_{BB\phi\phi}\ \cL (\mphi,\mu)
        + \sum_{\phi \phi^\prime} {\overline C}^M_{BB\phi \phi^\prime}
        \cL (\mphi,m_{\phi^\prime}) \bigg\} \nonumber\\
% M_+ M_+ insertion
     && - \frac{1}{(4 \pi f)} \sum_{i,j} C^{PQ}_{B_{ij}} m_i m_j
\end{eqnarray}
The coefficients for the NLO and NNLO octet baryon masses are listed
in Table~\ref{t:BBAPQ} through Table~\ref{t:BmassInsertPQ}.  It is
straight forward to show that in the limit $\mj \rightarrow \mbar$ and
$\mr \rightarrow \ms$, these expressions reduce to the $\chi$PT
expressions listed in Appendix~\ref{ap:XPTmasses}.
%
%
%
%   PQXPT Tables
%
%
%
\begin{table}[p]
\caption{\label{t:VVVmasses}
The LO $\mathbf {240}\ (B)$ and $\mathbf {138}\ (T)$ baryon masses
composed only of valence quarks in PQ$\chi$PT.}
\begin{ruledtabular}
\begin{tabular}{c | c | c | c }
   & Octet Baryons    & &   Decuplet Baryons    \\
   & $M_B^{(1)}$ & & $M_T^{(1)}$ \\
   \hline
     $\MN$ & $2\mbar (\am +\bm) +4\mj \smb +2\mr \smb$
   & $\MD$ & $2\mbar\gm -4\mj\smt -2\mr\smt$ \\ \hline
     $\MS$ & $\mbar \left(\frac{5}{3}\am +\frac{2}{3}\bm \right)
             +\ms \left(\frac{1}{3}\am +\frac{4}{3}\bm \right)$
   & $\MSS$ & $\frac{4}{3}\mbar\gm +\frac{2}{3}\ms\gm
              -4\mj\smt -2\mr\smt$ 
   \\
   & $+4\mj \smb +2\mr \smb$ & \\
   \hline
     $\ML$ & $\mbar (\am +2\bm) +\ms\am +4\mj\smb +2\mr\smb$
   & $\MXS$& $\frac{2}{3}\mbar\gm +\frac{4}{3}\ms\gm
             -4\mj\smt -2\mr\smt$ \\ \hline
     $\MX$ & $\mbar \left(\frac{1}{3}\am +\frac{4}{3}\bm \right)
             +\ms \left(\frac{5}{3}\am +\frac{2}{3}\bm \right)$
   & $\MO$ & $2\ms\gm -4\mj\smt -2\mr\smt$
   \\
   & $+4\mj \smb +2\mr \smb$ & & \\
\end{tabular}
\end{ruledtabular}
\end{table} 

\begin{table}[p]
\caption{\label{t:SGTmasses}
The LO masses of the $\mathbf {138}\ (T)$ baryons composed of $2$ valence quarks and $1$ sea
or ghost quark in PQ$\chi$PT.}
\begin{ruledtabular}
\begin{tabular}{c | c | c | c }
        & $\mathbf {138}$ Baryons & & $\mathbf {138}$ Baryons     \\
        & with one sea quark  & & with one ghost quark \\
        \hline
          $\MsSSj$ & $\frac{4}{3}\mbar \gm +\frac{2}{3}\mj (\gm
                     -6\smt) -2\mr\smt$
                  & $\MgSSu$ & $\MD$
        \\ \hline
          $\MsSSr$ & $\frac{4}{3}\mbar \gm -4\mj\smt +\frac{2}{3}\mr(\gm-3\smt)$
        & $\MgSSs$ & $\MSS$
        \\ \hline
          $\MsXSj$ & $\frac{2}{3}(\mbar +\ms)\gm
          +\frac{2}{3}\mj(\gm-6\smt) -2\mr\smt$
        & $\MgXSu$ & $\MSS$
        \\ \hline
          $\MsXSr$ & $\frac{2}{3}(\mbar +\ms)\gm -4\mj\smt
                     +\frac{2}{3}\mr(\gm-3\smt)$
        & $\MgXSs$ & $\MXS$
        \\ \hline
          $\MsOmj$ & $\frac{4}{3}\ms \gm +\frac{2}{3}\mj(\gm-6\smt)
                     -2\mr\smt$
        & $\MgOmu$ & $\MXS$
        \\ \hline
          $\MsOmr$ & $\frac{4}{3}\ms \gm -4\mj\smt
                     +\frac{2}{3}\mr(\gm-3\smt)$
        & $\MgOms$ & $\MO$
        \\
\end{tabular}
\end{ruledtabular}
\end{table}

\begin{table}[p]
\caption{\label{t:SGBmasses}
The LO masses of the $\mathbf {240}\ (B)$ baryons composed $2$ valence quarks and $1$
sea or ghost quark in PQ$\chi$PT.}
\begin{ruledtabular}
\begin{tabular}{c | c | c | c }
   & $\mathbf {240}$ Baryons & & $\mathbf {240}$ Baryons     \\
   & with one sea quark  & & with one ghost quark \\
   \hline
     $\MsSj$ & $\mbar \left(\frac{5}{3}\am +\frac{2}{3}\bm \right) +2\mr \smb$
   & $\MgSu$ & $\MN$
   \\
   & $+\mj \left(\frac{1}{3}\am +\frac{4}{3}\bm +4\smb \right)$ & \\ \hline
     $\MsSr$ & $\mbar \left(\frac{5}{3}\am +\frac{2}{3}\bm
               \right) +4\mj\smb$ 
   & $\MgSs$ & $\MS$
   \\
   & $+\mr \left(\frac{1}{3}\am +\frac{4}{3}\bm +2\smb \right)$ & \\
   \hline
     $\MsLj$ & $\mbar(\am+2\bm) +\mj(\am+4\smb) +2\mr\smb$
   & $\MgLu$ & $\MN$ \\ \hline
     $\MsLr$ & $\mbar(\am+2\bm) +4\mj\smb +\mr(\am+2\smb)$
     & $\MgLs$ & $\ML$ \\
   \hline
     $\MsXsj$& $(\mbar +\ms)\left( \frac{5}{6}\am +\frac{2}{6}\bm
               \right) +2\mr \smb$
     & $\MgXsu$& $\mbar\left(\frac{7}{6}\am +\frac{5}{3}\bm\right)
                 +4\mj\smb$
   \\
   & $+\mj \left(\frac{1}{3}\am +\frac{4}{3}\bm +4\smb \right)$ 
   & & $+\ms\left(\frac{5}{6}\am+\frac{1}{3}\bm\right) +2\mr\smb$ \\ \hline
     $\MsXsr$& $(\mbar +\ms)\left( \frac{5}{6}\am +\frac{2}{6}\bm
               \right) +4\mj \smb$
   & $\MgXss$& $\mbar\left(\frac{5}{6}\am+\frac{1}{3}\bm\right) +4\mj\smb$
   \\
   & $+\mr\left(\frac{1}{3}\am +\frac{4}{3}\bm +2\smb \right)$ 
   & & $+\ms\left(\frac{7}{6}\am +\frac{5}{3}\bm\right) +2\mr\smb$ \\
   \hline
     $\MsXtj$& $\frac{1}{2}(\mbar +\ms)(\am+2\bm) +\mj(\am+4\smb)$
   & $\MgXtu$& $\mbar\left(\frac{3}{2}\am+\bm\right) +4\mj\smb$
   \\
   & $+2\mr\smb$ 
   & & $+\ms\left(\frac{1}{2}\am+\bm\right) +2\mr\smb$ \\ \hline
     $\MsXtr$ & $\frac{1}{2}(\mbar +\ms)(\am+2\bm) +4\mj\smb$
   & $\MgXts$& $\mbar\left(\frac{1}{2}\am+\bm\right) +4\mj\smb$
   \\
   & $+\mr(\am+2\smb)$ 
   & & $+\ms\left(\frac{3}{2}\am+\bm\right) +2\mr\smb$ \\
   \hline
     $\MsOj$ & $\ms \left(\frac{5}{3}\am +\frac{2}{3}\bm \right) +2\mr\smb$
   & $\MgOu$ & $\MX$
   \\
   & $+\mj \left(\frac{1}{3}\am +\frac{4}{3}\bm +4\smb \right)$ & \\ \hline
     $\MsOr$ & $\ms \left(\frac{5}{3}\am +\frac{2}{3}\bm \right) +4\mj\smb$
   & $\MgOs$ & $4\mj\smb +2\ms(\am+\bm) +2\mr\smb$
   \\
   & $+\mr \left(\frac{1}{3}\am +\frac{4}{3}\bm +2\smb \right)$ & \\
\end{tabular}
\end{ruledtabular}
\end{table}

\begin{table}[p]
\caption{\label{t:BBAPQ}
  The $\mathbf {240}$-$\mathbf {240}$-Axial coupling coefficients in PQ$\chi$PT.
  The coefficients are grouped by the loop mesons with mass $\mphi$
  for each of the octet baryons.}
\begin{ruledtabular}
\begin{tabular}{c | c c c c }
   & & $\sum_\phi (C^{PQ}_{BB\phi})^2$ & & \\
   & $\pi$ & $K$ & & $\eta_s$ \\
   \hline
% Nucleon
   $N$ & $\frac{4}{3}D(3F-D)$ 
       & $0$
       &
       & $0$ \\ \hline
% Sigma
   $\S$ & $\frac{2}{3}(3F^2 -D^2)$ 
        & $-\frac{2}{3}(D^2 -6DF +3F^2)$ 
        &
        & $0$ \\ \hline
% Lambda
   $\L$ & $-\frac{2}{9}(D^2 -12DF +9F^2)$ 
        & $\frac{2}{9}(-5D^2 +6DF +9F^2)$ 
        &
        & $0$ \\ \hline
% Cascade
   $\X$ & $0$ 
        & $-\frac{2}{3}(D^2 -6DF +3F^2)$ 
        &
        & $\frac{2}{3}(3F^2 -D^2)$ \\
   \hline \hline \\
%   & & $\sum_\phi (C^{PQ}_{BB\phi})^2$ & & \\
   & $ju$ 
   & $ru$ 
   & $js$ 
   & $rs$ \\
   \hline
% Nucleon
   $N$ & $\frac{2}{3}(5D^2 -6DF+9F^2)$ 
       & $\frac{1}{3}(5D^2 -6DF +9F^2)$ 
       & $0$ 
       & $0$ \\ \hline
% Sigma
   $\S$ & $\frac{4}{3}(D^2 +3F^2)$ 
        & $\frac{2}{3}(D^2 +3F^2)$ 
        & $2(D-F)^2$ 
        & $(D-F)^2$ \\ \hline
% Lambda
   $\L$ & $\frac{4}{9}(7D^2 -12DF +9F^2)$ 
        & $\frac{2}{9}(7D^2 -12DF +9F^2)$ 
        & $\frac{2}{9}(D+3F)^2$ 
        & $\frac{1}{9}(D+3F)^2$ \\ \hline
% Cascade
   $\X$ & $2(D-F)^2$ 
        & $(D-F)^2$ 
        & $\frac{4}{3}(D^2 +3F^2)$ 
        & $\frac{2}{3}(D^2 +3F^2)$ \\
\end{tabular}
\end{ruledtabular}
\end{table}

\begin{table}[p]
\caption{\label{t:BBAhairpins}
  The coefficients for meson loops containing hairpin insertions with
an internal $\mathbf {240}$-baryon.}
\begin{ruledtabular}
\begin{tabular}{c | c c c}
   & & $\sum_{\phi \phi^\prime}{\overline C}_{B\phi \phi^\prime}$  & \\
   & $\eta_u \eta_u$ 
   & $\eta_u \eta_s$ 
   & $\eta_s \eta_s$ \\
   \hline
% Nucleon
   $N$ & $(D-3F)^2$ 
       & $0$  
       & $0$ \\ \hline
% Sigma
   $\S$ & $4F^2$ 
        & $4F(F-D)$ 
        & $(D-F)^2$ \\ \hline
% Lambda
   $\L$ & $\frac{4}{9}(2D-3F)^2$ 
        & $\frac{4}{9}(9F^2 -3DF -2D^2)$
        & $\frac{1}{9}(D+3F)^2$ \\ \hline
% Cascade
   $\X$ & $(D-F)^2$ 
        & $4F(F-D)$ 
        & $4F^2$ \\
\end{tabular}
\end{ruledtabular}
\end{table}

\begin{table}[p]
\caption{\label{t:TBAPQ}
  The $\mathbf {240}$-$\mathbf {138}$-Axial coupling coefficients and
  the coefficients for meson loops containing hairpin insertions with
  an internal $\mathbf {138}$-baryon in PQ$\chi$PT.  The coefficients
  are grouped by the loop mesons with mass $\mphi$ and by the
  participating $\eta$ fields in the hairpin interaction, respectively.}
\begin{ruledtabular}
\begin{tabular}{c | c c c c c c c | c c c}
   & & & & $\sum_\phi (C^{PQ}_{TB\phi})^2$ & & & & &
           $\sum_{\phi,\phi^\prime}{\overline C}_{T\phi \phi^\prime}$ \\
   & $\pi$ 
   & $K$ 
   & $\eta_s$ 
   & $ju$ 
   & $ru$ 
   & $js$ 
   & $rs$ 
   & $\eta_u \eta_u$
   & $\eta_u \eta_s$ 
   & $\eta_s \eta_s$ \\
   \hline
% Nucleon
   $N$ & $\frac{2}{3}$ 
       & $0$ 
       & $0$ 
       & $\frac{2}{3}$ 
       & $\frac{1}{3}$
       & $0$ 
       & $0$ 
       & $0$ 
       & $0$ 
       & $0$ \\ \hline
% Sigma
   $\S$ & $\frac{1}{9}$ 
        & $\frac{5}{9}$ 
        & $0$ 
        & $\frac{2}{9}$ 
        & $\frac{1}{9}$ 
        & $\frac{4}{9}$ 
        & $\frac{2}{9}$ 
        & $\frac{2}{9}$ 
        & $-\frac{4}{9}$ 
        & $\frac{2}{9}$ \\ \hline
% Lambda
   $\L$ & $\frac{1}{3}$ 
        & $\frac{1}{3}$ 
        & $0$ 
        & $\frac{2}{3}$ 
        & $\frac{1}{3}$
        & $0$ 
        & $0$  
        & $0$ 
        & $0$ 
        & $0$ \\ \hline
% Cascade
   $\X$ & $0$ 
        & $\frac{5}{9}$ 
        & $\frac{1}{9}$ 
        & $\frac{4}{9}$ 
        & $\frac{2}{9}$ 
        & $\frac{2}{9}$ 
        & $\frac{1}{9}$ 
        & $\frac{2}{9}$ 
        & $-\frac{4}{9}$ 
        & $\frac{2}{9}$ \\
\end{tabular}
\end{ruledtabular}
\end{table}

\begin{table}[p]
\caption{\label{t:BBAAPQ}
  The $C^{A,PQ}_{BB\phi\phi}$ and $ C^{vA,PQ}_{BB\phi\phi}$
  coefficients in PQ$\chi$PT.  The coefficients are grouped by the
  loop mesons with mass $\mphi$.}
\begin{ruledtabular}
\begin{tabular}{c | c c c c c c c c c c}
   & & & & & $\sum_\phi C^{A,PQ}_{BB\phi\phi}$ and & $\sum_\phi
               C^{vA,PQ}_{BB\phi\phi}$\\
   & $\pi$ 
   & $K$ 
   & $\eta_s$ 
   & $ju$ 
   & $ru$ 
   & $js$ 
   & $rs$ 
   & $\eta_j$ 
   & $jr$ 
   & $\eta_r$ \\
   \hline
% Nucleon
   $N$ & $-\frac{1}{2}b^A_3$ 
       & $0$ 
       & $0$ 
       & $2b^A_1 +2b^A_2$ 
       & $b^A_1 +b^A_2$ 
       & $0$ 
       & $0$ 
       & $4b^A_4$ 
       & $4b^A_4$ 
       & $b^A_4$ \\ \hline
% Sigma
   $\S$ & $\frac{1}{6}b^A_3$ 
        & $-\frac{2}{3}b^A_3$ & $0$ 
        & $\frac{2}{3}b^A_1 +\frac{5}{3}b^A_2$ 
        & $\frac{1}{3}b^A_1 +\frac{5}{6}b^A_2$ 
        & $\frac{4}{3}b^A_1 +\frac{1}{3}b^A_2$ 
        & $\frac{2}{3}b^A_1 +\frac{1}{6}b^A_2$ 
        & $4b^A_4$ 
        & $4b^A_4$ 
        & $b^A_4$ \\ \hline
% Lambda
   $\L$ & $-\frac{1}{2}b^A_3$ 
        & $0$ 
        & $0$ 
        & $2b^A_1 +b^A_2$ 
        & $b^A_1 +\frac{1}{2}b^A_2$ 
        & $b^A_2$ 
        & $\frac{1}{2}b^A_2$ 
        & $4b^A_4$ 
        & $4b^A_4$ 
        & $b^A_4$ \\ \hline
% Cascade
   $\X$ & $0$ 
        & $-\frac{2}{3}b^A_3$ 
        & $\frac{1}{6}b^A_3$ 
        & $\frac{4}{3}b^A_1 +\frac{1}{3}b^A_2$ 
        & $\frac{2}{3}b^A_1 + \frac{1}{6}b^A_2$ 
        & $\frac{2}{3}b^A_1 +\frac{5}{3}b^A_2$ 
        & $\frac{1}{3}b^A_1 +\frac{5}{6}b^A_2$ 
        & $4b^A_4$ & $4b^A_4$ 
        & $b^A_4$ \\
\end{tabular}
\end{ruledtabular}
\end{table}

\begin{table}[p]
\caption{\label{t:BBAAhairpin}
  The ${\overline C}^{A,PQ}_{BB\phi\phi}$ and ${\overline
    C}^{vA,PQ}_{BB\phi\phi}$ coefficients in PQ$\chi$PT.  The
  coefficients are grouped by the $\eta$ fields participating in the
  hairpin interaction.}
\begin{ruledtabular}
\begin{tabular}{c | c c c c c }
   & & $\sum_{\phi,\phi^\prime} {\overline C}^A_{BB\phi\phi^\prime}$ 
        & and $\sum_{\phi,\phi^\prime} {\overline C}^{vA}_{BB\phi\phi^\prime}$\\
   & $\eta_u \eta_u$ 
   & $\eta_u \eta_s$ 
   & $\eta_s \eta_s$ 
   & $\eta_j \eta_j$ 
   & $\eta_r \eta_r$ \\
   \hline
% Nucleon
   $N$ & $b^A_1 +b^A_2 +b^A_3$ 
       & $0$ 
       & $0$ 
       & $2b^A_4$ 
       & $b^A_4$ \\ \hline
% Sigma
   $\S$ & $\frac{1}{3}b^A_1 +\frac{5}{6}b^A_2 +\frac{1}{6}b^A_3$ 
        & $\frac{5}{6}b^A_3$ 
        & $\frac{2}{3}b^A_1 +\frac{1}{6}b^A_2$ 
        & $2b^A_4$ 
        & $b^A_4$ \\ \hline
% Lambda
   $\L$ & $b^A_1 +\frac{1}{2}b^A_2 +\frac{1}{2}b^A_3$ 
        & $\frac{1}{2}b^A_3$ 
        & $\frac{1}{2}b^A_2$ 
        & $2b^A_4$ & $b^A_4$ \\ \hline
% Cascade
   $\X$ & $\frac{2}{3}b^A_1 +\frac{1}{6}b^A_2$ 
        & $\frac{5}{6}b^A_3$ 
        & $\frac{1}{3}b^A_1 +\frac{5}{6}b^A_2 +\frac{1}{6}b^A_3$ 
        & $2b^A_4$ 
        & $b^A_4$ \\
\end{tabular}
\end{ruledtabular}
\end{table}

\begin{table}[p]
\caption{\label{t:MBBphiphiPQ}
  The $C^{M,PQ}_{BB\phi\phi}$ coefficients in PQ$\chi$PT.  The
  coefficients are grouped by the loop meson with mass $\mphi$.}
\begin{ruledtabular}
\begin{tabular}{c | c c c c }
   & & $\sum_\phi C^{M,PQ}_{BB\phi\phi}$ & & \\
   & $ju$ 
   & $ru$ 
   & $js$ \\
   \hline
% Nucleon
   $N$ & $4(\mbar +\mj)(\am +\bm)$ 
       & $2(\mbar +\mr)(\am +\bm)$ 
       & $0$ \\ \hline
% Sigma
   $\S$ & $(\mbar +\mj)\left( \frac{10}{3}\am +\frac{4}{3}\bm \right)$
        & $(\mbar +\mr)\left( \frac{5}{3}\am +\frac{2}{3}\bm \right)$ 
        & $(\mj +m_s) \left( \frac{2}{3}\am +\frac{8}{3}\bm \right)$ \\ \hline
% Lambda
   $\L$ & $(\mbar +\mj)(2\am +4\bm)$ 
        & $(\mbar +\mr)(\am +2\bm)$ 
        & $2(\mj +m_s)\am$ \\ \hline
% Cascade
   $\X$ & $(\mbar +\mj)\left( \frac{2}{3}\am +\frac{8}{3}\bm \right)$ 
        & $(\mbar +\mr) \left( \frac{1}{3}\am +\frac{4}{3}\bm \right)$
        & $(\mj +m_s)\left( \frac{10}{3}\am +\frac{4}{3}\bm \right)$ \\
% Part Deux
   \hline \hline \\
%   & & $\sum_\phi C^{M,PQ}_{BB\phi\phi}$ & & \\
   & $rs$ 
   & $\eta_j$ 
   & $jr$ 
   & $\eta_r$ \\
   \hline
% Nucleon
   $N$ & $0$ 
       & $16\mj \smb$ 
       & $8(\mj +\mr)\smb$ 
       & $4\mr\smb$\\ \hline
% Sigma
   $\S$ & $(m_s +\mr) \left( \frac{1}{3}\am +\frac{4}{3}\bm \right)$ 
        & $16\mj \smb$ 
        & $8(\mj +\mr)\smb$ 
        & $4\mr\smb$ \\ \hline
% Lambda
   $\L$ & $(m_s +\mr)\am $ 
        & $16\mj \smb$ 
        & $8(\mj +\mr)\smb$ 
        & $4\mr\smb$ \\ \hline
% Cascade
   $\X$ & $(\mr +m_s)\left( \frac{5}{3}\am +\frac{2}{3}\bm \right)$ 
        & $16\mj \smb$ 
        & $8(\mj +\mr)\smb$ 
        & $4\mr\smb$ \\
\end{tabular}
\end{ruledtabular}
\end{table}

%%%%%%%% Hairpins for M+ 2nd order
\begin{table}[p]
\caption{\label{t:MMhairpin}
  The ${\overline C}^{M}_{BB\phi\phi^\prime}$ coefficients in
  PQ$\chi$PT.  The coefficients are grouped by the $\eta$ fields
  participating in the hairpin interaction.}
\begin{ruledtabular}
\begin{tabular}{c | c c c c }
   & & $\sum_{\phi,\phi^\prime} {\overline C}^{M}_{BB\phi\phi^\prime}$ \\
   & $\eta_u \eta_u$ 
   & $\eta_s \eta_s$ 
   & $\eta_j \eta_j$ 
   & $\eta_r \eta_r$ \\
   \hline
% Nucleon
   $N$ & $4\mbar (\am +\bm)$ 
       & $0$ 
       & $8\mj \smb$ 
       & $4\mr \smb$ \\ \hline
% Sigma
   $\S$ & $\mbar \left( \frac{10}{3}\am + \frac{4}{3}\bm \right)$
        & $m_s \left( \frac{2}{3}\am +\frac{8}{3}\bm \right)$ 
        & $8\mj \smb$ 
        & $4\mr \smb$ \\ \hline
% Lambda
   $\L$ & $\mbar (2\am +4\bm)$ 
        & $2 m_s \am$ 
        & $8\mj \smb$ 
        & $4\mr \smb$ \\ \hline
% Cascade
   $\X$ & $\mbar \left( \frac{2}{3}\am + \frac{8}{3}\bm \right)$ 
        & $m_s \left( \frac{10}{3}\am +\frac{4}{3}\bm \right)$ 
        & $8\mj \smb$ 
        & $4\mr \smb$ \\
\end{tabular}
\end{ruledtabular}
\end{table}

%%%%%%%%%%%%% Internal Decuplet and Dec Wavefunction
\begin{table}[p]
\caption{\label{t:TmassPQ}
  The coefficients for internal $\mathbf {138}$-baryon mass insertions
  and the wavefunction corrections.  The coefficients are grouped by
  the loop mesons with mass $\mphi$ for each octet baryon.}
\begin{ruledtabular}
\begin{tabular}{c | c c c c }
   & & $\sum_{j,\phi}M^{(1)}_{T_j}(C^{PQ}_{TT\phi})^2\ \ +$ &
         $M^{(1)}_{B_i}\sum_\phi (C^{PQ}_{TT\phi})^2$ \\
   & $\pi$
   & $K$
   & $\eta_s$ \\
   \hline
% Nucleon
   $N$ & $\frac{2}{3}(\MN +\MD)$
       & $0$
       & $0$ \\ \hline
% Sigma
   $\S$ & $\frac{1}{9}(\MS +\MSS)$
        & $\frac{1}{9}(5\MS +4\MD +\MXS)$
        & $0$ \\ \hline
% Lambda
   $\L$ & $\frac{1}{3}(\ML +\MSS)$
        & $\frac{1}{3}(\ML +\MXS)$
        & $0$ \\ \hline
% Cascade
   $\X$ & $0$
        & $\frac{1}{9}(5\MX +4\MO +\MSS)$
        & $\frac{1}{9}(\MX +\MXS)$ \\
   \hline \hline \\
%   & & $\sum_{j,\phi}M^{(1)}_{T_j}(C^{PQ}_{TT\phi})^2\ \ +$ &
%         $M^{(1)}_{B_i}\sum_\phi (C^{PQ}_{TT\phi})^2$ \\
   & $ju$
   & $ru$
   & $js$
   & $rs$ \\
   \hline
% Nucleon
   $N$ & $\frac{2}{3}(\MN +\MsSSj)$
       & $\frac{1}{3}(\MN +\MsSSr)$
       & $0$
       & $0$ \\ \hline
% Sigma
   $\S$ & $\frac{2}{9}(\MS +\MsXSj)$
        & $\frac{1}{9}(\MS +\MsXSr)$
        & $\frac{4}{9}(\MS +\MsSSj)$
        & $\frac{2}{9}(\MS +\MsSSr)$ \\ \hline
% Lambda
   $\L$ & $\frac{2}{3}(\ML +\MsXSj)$
        & $\frac{1}{3}(\ML +\MsXSr)$
        & $0$
        & $0$ \\ \hline
% Cascade
   $\X$ & $\frac{4}{9}(\MX +\MsOj)$
        & $\frac{2}{9}(\MX +\MsOr)$
        & $\frac{2}{9}(\MX +\MsXSj)$
        & $\frac{1}{9}(\MX +\MsXSr)$ \\
\end{tabular}
\end{ruledtabular}
\end{table}
%%%%%%%%%%%%%  Internal Dec Hairpins
\begin{table}[p]
\caption{\label{t:TmassHairpins}
  The coefficients for hairpin interactions with internal $\mathbf
  {138}$-baryon mass insertions and wavefunction corrections.  The
  coefficients are grouped according to the $\eta$ fields
  participating in the hairpin interaction.}
\begin{ruledtabular}
\begin{tabular}{c | c c c }
   & $\sum_{j,\phi,\phi^\prime} M^{(1)}_{T_j} {\overline
     C}_{T\phi\phi^\prime}\ \ +$
   & $M^{(1)}_{B_i} \sum_{\phi,\phi^\prime} {\overline
     C}_{T\phi\phi^\prime}$ \\
   & $\eta_u \eta_u$
   & $\eta_u \eta_s$
   & $\eta_s \eta_s$ \\
   \hline
% Nucleon
   $N$ & $0$
       & $0$
       & $0$ \\ \hline
% Sigma
   $\S$ & $\frac{2}{9}(\MS +\MSS)$
        & $-\frac{4}{9}(\MS +\MSS)$
        & $\frac{2}{9}(\MS +\MSS)$ \\ \hline
% Lambda
   $\L$ & $0$
        & $0$
        & $0$ \\ \hline
% Cascade
   $\X$ & $\frac{2}{9}(\MX +\MXS)$
        & $-\frac{4}{9}(\MX +\MXS)$
        & $\frac{2}{9}(\MX +\MXS)$ \\
\end{tabular}
\end{ruledtabular}
\end{table}
%%%%%%%%%%%%%% M+M+
\begin{table}[p]
\caption{\label{t:MMPQ}
  The $C^{PQ}_{B_{ij}}$ coefficients in PQ$\chi$PT.  The coefficients
  are grouped by products of quark masses.}
\begin{ruledtabular}
\begin{tabular}{c | c c c c c }
   & & & $\sum_{i,j} C^{PQ}_{B_{ij}}$ \\
   & $\mbar^2$ 
   & $\mbar \mj$ 
   & $\mj^2$ 
   & $\mbar m_s$ 
   & $\mbar \mr$
       \\
       \hline
% Nucleon
   $N$ & $b^M_1 +b^M_2 +b^M_3$ 
       & $2b^M_5 +2b^M_6$ 
       & $2b^M_4 + 4b^M_7$ 
       & $0$ 
       & $b^M_5 +b^M_6$ \\ \hline
% Sigma
   $\S$ & $\frac{1}{3}b^M_1 +\frac{5}{6}b^M_2 +\frac{1}{6}b^M_3$ 
        & $\frac{2}{3}b^M_5 +\frac{5}{3}b^M_6$ 
        & $2b^M_4 +4b^M_7$ 
        & $\frac{5}{6}b^M_3$ 
        & $\frac{1}{3}b^M_5 +\frac{5}{6}b^M_6$ \\ \hline
% Lambda
   $\L$ & $b^M_1 +\frac{1}{2}b^M_2 +\frac{1}{2}b^M_3$ 
        & $2b^M_5 +b^M_6$ 
        & $2b^M_4 +4b^M_7$ 
        & $\frac{1}{2}b^M_3$ & $b^M_5 +\frac{1}{2}b^M_6$ \\ \hline
% Cascade
   $\X$ & $\frac{2}{3}b^M_1 +\frac{1}{6}b^M_2$ 
        & $\frac{4}{3}b^M_5 +\frac{1}{3}b^M_6$ 
        & $2b^M_4 +4b^M_7$ 
        & $\frac{5}{6}b^M_3$ 
        & $\frac{2}{3}b^M_5 +\frac{1}{6}b^M_6$ \\
   \hline \hline \\
%   & & & $\sum_{i,j} C^{PQ}_{B_{ij}}$ \\
   & $\mj m_s$ 
   & $\mj \mr$ 
   & $m_s \mr$ 
   & $m_s^2$ 
   & $\mr^2$ \\
   \hline
% Nucleon
   $N$ & $0$ 
       & $4b^M_7$ 
       & $0$ 
       & $0$ 
       & $b^M_4 +b^M_7$ \\ \hline
% Sigma
   $\S$ & $\frac{4}{3}b^M_5 +\frac{1}{3}b^M_6$ 
        & $4b^M_7$ 
        & $\frac{2}{3}b^M_5 +\frac{1}{6}b^M_6$ 
        & $\frac{2}{3}b^M_1 +\frac{1}{6}b^M_2$ 
        & $b^M_4 +b^M_7$ \\ \hline
% Lambda
   $\L$ & $b^M_6$ 
        & $4b^M_7$
        & $\frac{1}{2}b^M_6$ 
        & $\frac{1}{2}b^M_2$ 
        & $b^M_4 +b^M_7$ \\ \hline
% Cascade
   $\X$ & $\frac{2}{3}b^M_5 +\frac{5}{3}b^M_6$ 
        & $4b^M_7$ 
        & $\frac{1}{3}b^M_5 +\frac{5}{6}b^M_6$ 
        & $\frac{1}{3}b^M_1 +\frac{5}{6}b^M_2 +\frac{1}{6}b^M_3$ 
        & $b^M_4 +b^M_7$ \\
\end{tabular}
\end{ruledtabular}
\end{table}

%%%%%%%%%%%%%% Internal Mass Insertions
\begin{turnpage}
\begingroup
\squeezetable
\begin{table}[p]
\caption{\label{t:BmassInsertPQ}
  The coefficients for internal $\mathbf {240}$-baryon mass insertion
  and the wavefunction corrections.  The coefficients are grouped
  according to the loop mesons with mass $\mphi$ for each octet baryon.}
\begin{ruledtabular}
\begin{tabular}{c| c c c }
   & & $M^{(1)}_{B_i} \sum_\phi (C^{PQ}_{BB\phi})^2
        -\sum_{j,\phi} M^{(1)}_{B_j}(C^{PQ}_{BB\phi})^2$ \\
   & $\pi$ 
   & $K$ 
   & $\eta_s$ \\
   \hline
% Nucleon
   $N$ & $0$ 
       & $0$ 
       & $0$ \\
   \hline
% Sigma
   $\S$ & $(D-F)^2 \MgXsu +\frac{1}{3}(D+3F)^2 \MgXtu$ 
        & $(D-F)^2 \MN +\frac{1}{2}(D-F)^2 \MgXsu
           +\frac{1}{6}(D+3F)^2 \MgXtu$ 
        & $0$
        \\
        & $ -\frac{2}{3}(D^2+6F^2) \MS -\frac{2}{3}D^2 \ML$ 
        & $ -\frac{2}{3}(D^2-6DF+3F^2)\MS -(D+F)^2 \MX$ \\ 
   \hline
% Lambda
   $\L$ & $3(D-F)^2 \MgXsu +\frac{1}{9}(D+3F)^2 \MgXtu$
        & $\frac{3}{2}(D-F)^2 \MgXss +\frac{1}{18}(D+3F)^2\MgXts
           -\frac{1}{3}(D-3F)^2 \MX$ 
        & $0$
        \\
        & $-\frac{2}{9}(5D^2-24DF+18F^2) \ML -2D^2 \MS$ 
        & $-\frac{2}{9}(5D^2-6DF-9F^2)\ML -\frac{1}{9}(D+3F)^2 \MN$ \\ 
   \hline
% Cascade
   $\X$ & $0$
        & $\frac{1}{3}(D+3F)^2 \MgXtu +(D-F)^2 \Big( \MgOs +\MgXsu \Big)
           -\frac{1}{6}(D-3F)^2 \ML$
        & $\frac{1}{2}(D-F)^2 \MgXss +\frac{1}{6}(D+3F)^2 \MgXts$ 
        \\
        & & $-\frac{2}{3}(D^2 -6DF+3F^2)\MX -\frac{3}{2}(D+F)^2 \MS  $
        & $-\frac{2}{3}(D^2+3F^2) \MX$ \\
   \hline \hline \\
% Part Deux
%   & & $\sum_{j,\phi} M^{(1)}_{B_j}(C^{PQ}_{BB\phi})^2$ \\
   & $ju$
   & $ru$ \\
   \hline
% Nucleon
   $N$ & $\frac{2}{3}(5D^2 -6DF +9F^2) \MN$
       & $\frac{1}{3}(5D^2 -6DF +9F^2) \MN$ 
   \\
       & $-3(D-F)^2 \MsSj -\frac{1}{3}(D+3F)^2 \MsLj$
       & $-\frac{3}{2}(D-F)^2 \MsSr -\frac{1}{6}(D+3F)^2 \MsLr$\\
   \hline
% Sigma
   $\S$ & $\frac{4}{3}(D^2 +3F^2) \MS$ 
        & $\frac{2}{3}(D^2 +3F^2) \MS$ 
   \\
        & $-(D-F)^2 \MsXsj -\frac{1}{3}(D+3F)^2 \MsXtj$
        & $-\frac{1}{2}(D-F)^2 \MsXsr -\frac{1}{6}(D+3F)^2 \MsXtr$ \\
   \hline
% Lambda
   $\L$ & $\frac{4}{9}(7D^2 -12DF +9F^2)\ML$
        & $\frac{2}{9}(7D^2 -12DF +9F^2)\ML$ 
   \\
        & $-3(D-F)^2 \MsXsj -\frac{1}{9}(D+3F)^2 \MsXtj$
        & $-\frac{3}{2} (D-F)^2 \MsXsr -\frac{1}{18}(D+3F)^2 \MsXtr$ \\
   \hline
% Cascade
   $\X$ & $2(D-F)^2 (\MX -\MsOj)$
        & $(D-F)^2 (\MX -\MsOr)$ \\
   \hline \hline \\
% Part Trois
%   & & $\sum_{j,\phi} M^{(1)}_{B_j}(C^{PQ}_{BB\phi})^2$ \\
   & & $js$ 
   &   $rs$ \\
   \hline
% Nucleon
   $N$ & & $0$ 
       &  $0$ \\
   \hline
% Sigma
   $\S$ & & $2(D-F)^2 (\MS -\MsSj)$
        & $(D-F)^2 (\MS -\MsSr)$ \\
   \hline
% Lambda
   $\L$ & & $\frac{2}{9}(D+3F)^2 (\ML -\MsLj)$
        & $\frac{1}{9}(D+3F)^2 (\ML -\MsLr)$ \\
   \hline
% Cascade
   $\X$ & & $\frac{4}{3}(D^2+3F^2)\MX$
        & $\frac{2}{3}(D^2+3F^2)\MX$ 
   \\
        & & $-(D-F)^2 \MsXsj -\frac{1}{3}(D+3F)^2 \MsXtj$
          & $-\frac{1}{2} (D-F)^2 \MsXsr -\frac{1}{6}(D+3F)^2 \MsXtr$ \\
\end{tabular}
\end{ruledtabular}
\end{table}
\endgroup
\end{turnpage}

%%%%%%%%%  References  %%%%%%%%%
\pagebreak
\bibliography{biblio}

\end{document}